\documentclass[a4paper,11pt]{article}
\pdfoutput=1 

\usepackage{jheppub1} 

\usepackage[T1]{fontenc} 
\usepackage{empheq,mathtools}
\usepackage{cancel,slashed}
\usepackage{xcolor}
\usepackage{verbatim}
\usepackage{float}
\usepackage{subfigure}
\usepackage{framed}

\def\tsp{\hspace{0.083333em}}
\def\arXiv#1{\href{http://arxiv.org/abs/#1}{arXiv:#1}}
\def\arXiv#1#2{\href{http://arxiv.org/abs/#1}{arXiv:#1}}
\def\arXivid#1#2{\href{http://arxiv.org/abs/#1/#2}{#1/#2}}
\def\doi#1#2{\href{http://doi.org/#1}{#2}}

\title{\boldmath Holographic complexity of hyperbolic black holes}

\author{Yongao Wang and Jie Ren}

\affiliation{School of Physics, Sun Yat-sen University, Guangzhou, 510275, China}

\emailAdd{wangyao27@mail2.sysu.edu.cn, renjie7@mail.sysu.edu.cn}

\abstract{

We calculate the holographic complexity of a family of hyperbolic black holes in an Einstein-Maxwell-dilaton (EMD) system by applying the complexity=action (CA) conjecture. While people previously studied spherical black holes in the same system, we show that hyperbolic black holes have intriguing features. We confirm that the complexity expression mainly depends on the spacetime causal structure despite the rich thermodynamics. The nontrivial neutral limit that exists only for hyperbolic black holes enables us to analytically obtain the complexity growth rate during phase transitions. We find that the dilaton accelerates the growth rate of complexity, and the Lloyd bound can be violated. This is in contrast to the spherical case where the dilation slows the complexity growth rate down, and the Lloyd bound is always satisfied. As a special case, we study the holographic complexity growth rate of the neutral hairy black holes and find that the Lloyd bound is always violated.}

\arxivnumber{2304.10454}

\begin{document}

\maketitle

\flushbottom

\section{Introduction}
\label{sec:intro}
One of the most exciting observations in theoretical physics is the deep connection between gravity and information, which originates from the research of black hole entropy \cite{Bekenstein:1973ur}. After the proposal of AdS/CFT conjecture \cite{Maldacena:1997re,Witten:1998qj}, this connection is becoming more and more convincing and rises mountains of research interest. Ryu and Takayanagi \cite{Ryu:2006bv} then formulated a proposal that relates the entanglement entropy in the boundary theory to the areas of minimal surfaces in the anti-de Sitter (AdS) spacetime (see \cite{Nishioka:2009un} for a review). 
This proposal strongly suggests that spacetime can be regarded as an emergent phenomenon related to quantum entanglement. However, as the research progresses, it is noticed that the black hole horizon blocks our way to understand spacetime in terms of entanglement. One evidence is that the minimal surfaces in the Ryu-Takayanagi proposal are unable to probe the whole geometry behind the black hole horizon. Additionally, the volume behind the horizon keeps growing in late time while the entanglement has already been saturated \cite{Hartman:2013qma}. These facts implicitly suggest that entanglement entropy is not enough and bring another quantity from information theory called quantum complexity into our sight \cite{Susskind:2014moa,Susskind:2014rva,Brown:2015bva,Stanford:2014jda,Brown:2015lvg}. Briefly speaking, quantum complexity is usually defined as the least possible number of unitary transformations needed for constructing a target state from a given reference state. In chaotic systems, the complexity will be expected to grow linearly and be sensitive to perturbations. These properties inspire people to notice that certain geometry quantities about black holes also have such behavior and these similarities lead us to the conjectures of holographic complexity. 

There are several different proposals for the holographic description of complexity. The first is known as the complexity=volume (CV) conjecture, which identifies the complexity with the volume $\mathcal{V}$ of the maximum codimension-one surface anchored to the boundary, 
\begin{equation}
    \mathcal{C}_\text{V} = \frac{\max\left(\mathcal{V}\right)}{G_\text{N} L}
\end{equation}
where $G_\text{N}$ is Newton's constant and $L$ is the AdS length. The second proposal is called the complexity=action (CA) conjecture, which relates the complexity of the boundary theory  to the action evaluated in an exact bulk region with null boundaries known as the Wheeler-DeWitt (WDW) patch, 
\begin{equation}
    \mathcal{C}_\text{A} = \frac{S_\text{WDW}}{\pi}\,.
\end{equation}
Another proposal is called CV2.0, which states that the holographic complexity can be calculated by the spacetime volume of the WDW patch, 
\begin{equation}
\mathcal{C}_\text{SV} = \frac{\mathcal{V}_\text{WDW}}{G_\text{N} L^2}\,.
\end{equation}

Aside from these stated conjectures, recently there has been a series of works arguing that it is possible to construct an infinite family of gravitational observables that naturally include all the conjectures mentioned above \cite{Belin:2022xmt,Belin:2021bga}. All these different proposals have their own values that deserve further research. However, in this work, we will only focus on the CA conjecture and leave the possible generalization to future works.

In string theory, a scalar field called dilaton occurs in the low energy limit and couples to other gauge fields nontrivially; this coupling may change the spacetime structure and leave us with intriguing black hole solutions. As in simpler cases, the Einstein-Maxwell-dilaton (EMD) systems admit analytic solutions of charged dilaton black holes \cite{Garfinkle:1990qj}.\footnote{The scalar hair considered in this paper is neutral instead of charged. As for the case with charged scalar hair, see \cite{Auzzi:2022bfd}.} It can be considered as an extension of the Einstein-Maxwell theory with an exponential coupling between the scalar field and the Maxwell field. The EMD systems admit black brane solutions with hyperscaling violation and the complexity of these solutions was studied in \cite{Swingle:2017zcd,Alishahiha:2018tep}. Moreover, with a negative cosmological constant, the EMD systems have broad applications in the AdS/CFT correspondence. See \cite{Mahapatra:2018gig} for Einstein-dilaton systems. In most of the previous works on these black holes, authors focused on spherical or planar black hole solutions. The hyperbolic black holes deserve more attention and further exploration. 

We calculate the complexity from a class of EMD systems belonging to $\mathcal{N}=2$ supergravity \cite{Faedo:2015jqa}. Special cases of this system can be embedded in 11-dimensional supergravity. The action is \eqref{eq:action4} with \eqref{eq:potential4} below. The complexity of the spherical black holes was calculated in \cite{Goto:2018iay}. The hyperbolic counterpart has the following intriguing features:
\begin{itemize}

    \item This system has a nontrivial neutral limit in which an Einstein-scalar system is obtained while the scalar field is kept nontrivial \cite{Ren:2019lgw}. Only in the hyperbolic case can we obtain a neutral hairy black hole.\footnote{The planar black hole as a nontrivial neutral limit in which its IR geometry is a hyperscaling-violating geometry, while the UV is asymptotically AdS.}

    \item For the Einstein-scalar system, the scalar field condensates without being sourced. The analytic solution describing the spontaneous development of the scalar hair is extremely rare. An application of this system is to study the phase transition of R\'{e}nyi entropies \cite{Bai:2022obp}.

    \item Thermodynamics of hyperbolic black holes has richer properties. While the parameters $b$ and $c$ must be positive for spherical black holes, they can be both negative for hyperbolic black holes.
\end{itemize}

In this work, we investigate the holographic complexity of the hyperbolic charged dilaton black holes by applying the CA conjecture. By comparing our result with the studied spherical case, we find similarities as well as essential differences between these two results. The main conclusions are as follows:
\begin{itemize}
\item The expression of the complexity mainly depends on the causal structure of the spacetime, despite the rich thermodynamics of hyperbolic black holes.
\item By studying the neutral case, we provide an analytic example that the complexity growth rate undergoes phase transitions as we decrease the temperature.
\item We find that the dilaton will accelerate the complexity growth instead of slowing it down, which is opposite to the spherical case.
\item The Lloyd bound is violated in certain parameter space, in contrast to the spherical case in which the Lloyd bound is always satisfied.
\end{itemize}
As a special case, we obtain the holographic complexity growth rate of the neutral hairy black holes as a nontrivial neutral limit, and find that the Lloyd bound is always violated due to the accelerated effect of the dilaton.

This paper is organized as follows. In section~\ref{sec:EMD}, we review the solution of the EMD system. In section~\ref{sec:CAcal}, we calculate the complexity of the neutral and charged black holes by CA proposal in subsections \ref{sec:3.1} and \ref{sec:3.2} respectively. In section~\ref{sec:diss}, we summarize our results and compare them with the spherical and planar solutions, and we discuss the violation of the Lloyd bound. In appendix~\ref{sec:noh}, we discuss the number of horizons. In appendix~\ref{sec:ap}, we briefly calculate the complexity of the planar EMD black hole. In appendix~\ref{sec:hc5d}, we give the result for a higher-dimensional EMD system.

\section{The EMD system and its hyperbolic black hole solution}
\label{sec:EMD}
In this section, we review some basic properties of the EMD system to prepare for our further calculations. 

The action is
\begin{equation}
S=\int d^4x\sqrt{-g}\left(R-\frac14 e^{-\alpha\phi}F^2-\frac12(\partial\phi)^2-V(\phi)\right),
\label{eq:action4}
\end{equation}
where $F=dA$, the bulk spacetime is asymptotically AdS$_4$ with the potential of the dilaton field as
\begin{equation}
V(\phi)=-\frac{2}{(1+\alpha^2)^2L^2}\Bigl[\alpha^2(3\alpha^2-1)e^{-\phi/\alpha}
+8\alpha^2e^{(\alpha-1/\alpha)\phi/2}+(3-\alpha^2)e^{\alpha\phi}\Bigr],\label{eq:potential4}
\end{equation}
where $\alpha$ is a parameter, and the values of $\alpha=0$, $1/\sqrt{3}$, $1$, and $\sqrt{3}$ correspond to special cases of STU supergravity. This potential was found in \cite{Gao:2004tu} and was rediscovered many times. When $\phi\to 0$, $V(\phi)\to-6/L^2-(1/L^2)\phi^2+\cdots$, where the first term is the cosmological constant, and the second term gives the mass $m^2L^2=-2$.

A solution of the metric $g_{\mu\nu}$, gauge field $A_\mu$, and dilaton field $\phi$ is \cite{Gao:2004tu}
\begin{align}
& ds^2=-f(r)\tsp dt^2+\frac{1}{f(r)}\tsp dr^2+U(r)\tsp d\tsp\Sigma_{2,k}^2\,,\label{eq:sol4}\\
& A=2\sqrt{\frac{bc}{1+\alpha^2}}\left(\frac{1}{r_+}-\frac{1}{r}\right)dt\,,\label{eq:gauge}\\ 
& e^{\alpha\phi}=\left(1-\frac{b}{r}\right)^\frac{2\alpha^2}{1+\alpha^2},\label{eq:dilaton}
\end{align}
with
\begin{equation}
\begin{split}
f &=\left(k-\frac{c}{r}\right)\left(1-\frac{b}{r}\right)^\frac{1-\alpha^2}{1+\alpha^2}+\frac{r^2}{L^2}\left(1-\frac{b}{r}\right)^\frac{2\alpha^2}{1+\alpha^2},\\
U &=r^2\left(1-\frac{b}{r}\right)^\frac{2\alpha^2}{1+\alpha^2}.\label{eq:fsol4}
\end{split}
\end{equation}
The parameter $k$ can be set to $-1\,,0\,,1$ corresponding to hyperbolic, planar, and spherical cases, respectively.
The horizon of the black hole is determined by $f(r_+)=0$.
\begin{equation}\label{c_of_rh}
c=kr_++\frac{r_+^3}{L^2}\left(1-\frac{b}{r_+}\right)^\frac{3\alpha^2-1}{1+\alpha^2}.
\end{equation}
We can see crucial differences between spherical, planar, and hyperbolic black holes. Notice that charged black holes have $bc\neq 0$.
\begin{itemize}
    \item Spherical black holes ($k=1$). We always have $c>r_+>b>0$. This is the case studied in \cite{Goto:2018iay}.
    \item Planar black holes ($k=0$). The case $c=0$ does not give a black hole; however, it has physical meaning as the extremal limit of black holes.
    \item Hyperbolic black holes ($k=-1$). The parameter space for black holes is significantly larger. For charged black holes, we have (i) $r_+>b>0$, $c>0$; (ii) $b<0<r_+$, $c<0$. The case $c=0$ gives a hairy neutral hyperbolic black hole.
\end{itemize}
The temperature and the entropy of the black hole are given by
\begin{equation}\label{tands}
T=\frac{f'(r_+)}{4\pi},\qquad S=\frac{V_2}{4G_N}U(r_+)\,,
\end{equation}
where $V_2$ is the area of two-dimensional space $d\tsp\Sigma_{2,k}^2$. The mass of this black hole is given by\footnote{A rigorous derivation of the boundary stress tensor by holographic renormalization can be found in appendix A of \cite{Bai:2022obp}.} \cite{Hendi:2010gq}
\begin{equation}\label{massgen}
    M = \frac{V_2}{8\pi G_N}\left(c+k\frac{1-\alpha^2}{1+\alpha^2}b\right).
\end{equation}
For $k=1$, $V_2=4\pi$; for $k=0$ or $-1$, $V_2$ is divergent. We can consider the entropy density $s\equiv S/V_2$ and energy density $\varepsilon\equiv M/V_2$. The chemical potential $\mu$ and charge density $q_e$ are
\begin{equation}
\mu=2\sqrt{\frac{bc}{1+\alpha^2}}\,\frac{1}{r_+},\qquad q_e=2\sqrt{\frac{bc}{1+\alpha^2}}\,.
\end{equation}
It is straightforward to verify that the first law of thermodynamics is satisfied ($16\pi G_N=1$),
\begin{equation}
d\varepsilon=Tds+\mu dq_e\,.
\end{equation}

\begin{figure}[H]
    \subfigure[single horizon structure when $\alpha^2\geq \frac13$]{\includegraphics[width=0.45\textwidth]{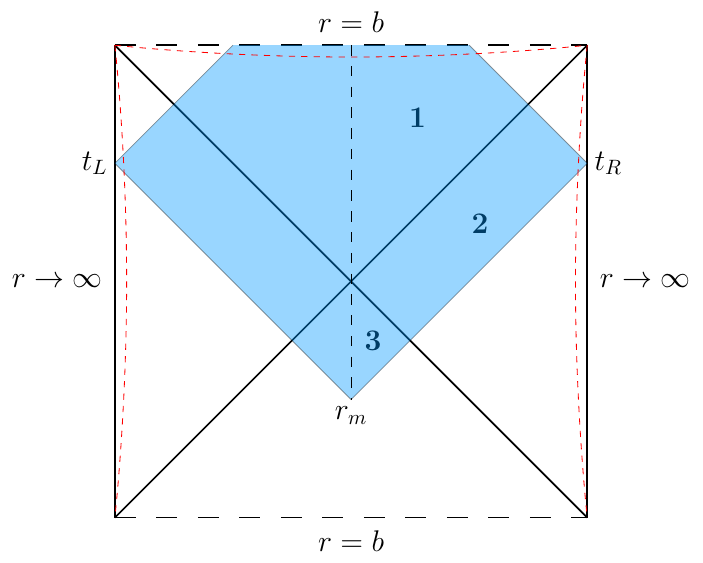}\label{single}}
    \subfigure[double horizon structure when $\alpha^2< \frac13$ ]{\centering \includegraphics[width=0.45\textwidth]{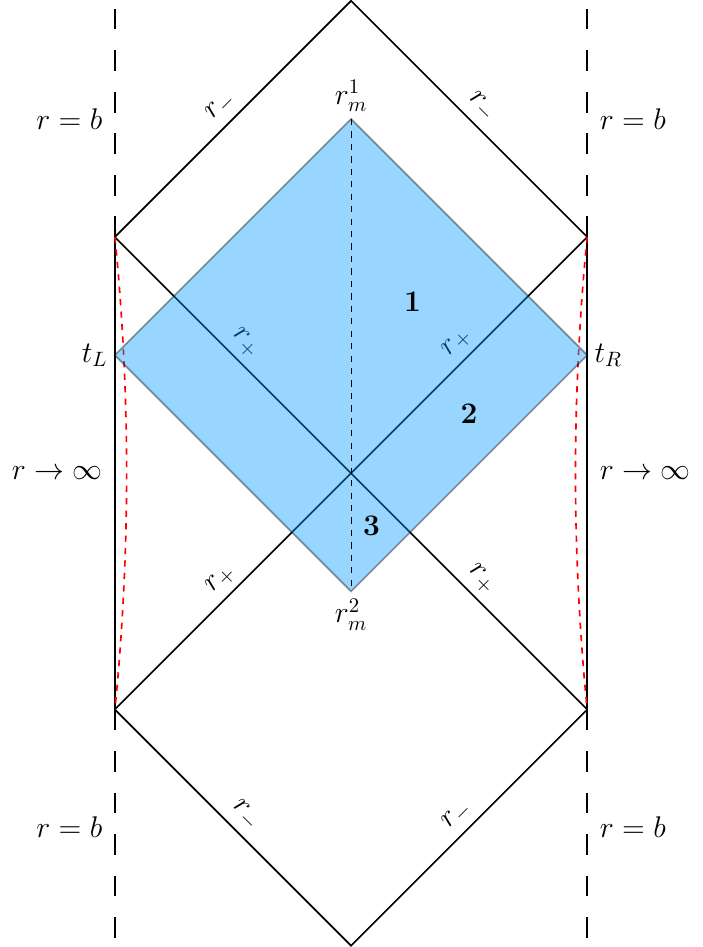}\label{double}}
    \caption{Causal structure of the black holes with $b>0$: (a) $\alpha^2\geq 1/3$, (b) $\alpha^2<1/3$.}
    \label{f1}
\end{figure}

In this work, we focus on the hyperbolic black hole solution $(k=-1)$. The solution has three other parameters $b$, $c$, and $\alpha$. The causal structure of this solution, which is extremely important in the computation of the CA proposal, depends on the value of these parameters. The parameter $\alpha$ controls the strength of the coupling of the scalar field to the gauge field. In the limit $\alpha\to0$ and $\alpha\to\infty$, the solution reduces to the Reissner-N\"{o}rdstrom-AdS (RN-AdS) black hole and the Schwarzschild-AdS (SAdS) black hole, respectively. Generally, the causal structure is determined by the solution of the equation $f(r)=0$. The causal structure of the spherical case has been introduced in \cite{Goto:2018iay}. The hyperbolic case is more complicated because the parameter $b$ can be negative here. In the charged case with $b>0$, for $\alpha^2\geq 1/3$, there is only one black hole horizon so that the causal structure is similar to the SAdS black hole as figure~\ref{single}. For $0<\alpha^2<1/3$, an additional horizon appears and the causal structure will be the type of RN-AdS black hole as figure~\ref{double}. With $b<0$ and $c<0$, there will always be two horizons for any $\alpha$. Hence the causal structure will also be shown by figure~\ref{double}, while the singularity is located at $r=0$ instead of $r=b$.

The solution \eqref{eq:sol4}--\eqref{eq:fsol4} has two neutral limits. When $b=0$, it is obvious that both the gauge field and dilaton field are eliminated. The solution reduces to the hyperbolic Schwarzchild-AdS black hole. When $c=0$, the dilaton field is kept, while the gauge field is eliminated. Under this limit, we obtain neutral hyperbolic black holes with scalar hair \cite{Ren:2019lgw}. The metric is
\begin{equation}
    ds^2 =-f(r)\tsp dt^2+\frac{1}{f(r)}\tsp dr^2+U(r)\tsp d\tsp\Sigma_{2}^2\,,
\end{equation}
where $d\tsp\Sigma_2^2=d\theta^2+\sinh^2\theta\tsp d\varphi^2$ is the hyperbolic $2$-space, and
\begin{equation}
\begin{split}
f &=-\left(1-\frac{b}{r}\right)^\frac{1-\alpha^2}{1+\alpha^2}+\frac{r^2}{L^2}\left(1-\frac{b}{r}\right)^\frac{2\alpha^2}{1+\alpha^2},\\
U &=r^2\left(1-\frac{b}{r}\right)^\frac{2\alpha^2}{1+\alpha^2},\qquad e^{\alpha\phi}=\left(1-\frac{b}{r}\right)^\frac{2\alpha^2}{1+\alpha^2}.\label{eq:fsol4-1}
\end{split}
\end{equation}
This neutral black hole solution can only be realized in the hyperbolic case.  With positive $b$, we find that the causal structure of the neutral solution is the same as the charged case as figure~\ref{f1}. The solution with negative $b$ can be obtained by the positive case by a transformation $b\to-b, \alpha\to1/\alpha$; namely, the black hole has one horizon when $\alpha^2\leq3$, two horizons when $\alpha^2>3$. A comprehensive analysis of the number of horizons can be found in appendix~\ref{sec:noh}. 


\section{Complexity growth from CA conjecture}\label{sec:CAcal}
\subsection{CA conjecture}
\label{sec:CA}
In this section, we compute the growth rate of the holographic complexity of the hyperbolic black holes by the CA conjecture, including both charged and neutral solutions. We begin with introducing some general ingredients needed for the calculation, which follows the method developed in \cite{Lehner:2016vdi}.

We first consider the case when spacetime has a single horizon as shown in figure~\ref{f1}(a).
Before a critical time $t_c$, the intersection between the WDW patch and the past singularity is a spacelike hypersurface. A useful result is that the WDW action is time independent before the critical time \cite{Carmi:2017jqz}. So we only need to study the period after critical time. After the critical time, the WDW patch will leave the past singularity and two null boundaries of the WDW patch will intersect with each other at a point $r=r_m$. We use the Eddington-Finkelstein coordinates given by
\begin{equation}
 v = t + r^* (r)\, ,\qquad u = t - r^* (r)\,, \qquad r^*(r) = \int\!\frac{dr}{f(r)}\,.
\end{equation}
At the critical time, we can write the null boundary as $u=u_0$ in the Eddington-Finkelstein coordinates. Then the critical time can be shown as
\begin{equation}
    \frac{t_\text{c}}{2}-r^*(\infty)=u_0=-r^*(r_m)\quad\Longrightarrow\quad t_c=2\left(r^*(\infty)-r^*(r_m)\right)\,.
\end{equation}
Similarly, the null boundary intersection $r_m$ satisfies a relation
\begin{equation}
    \frac{t}{2}-r^*_{\infty}+r^*(r_m) = 0\,,
\end{equation}
from which we can obtain the implicit time dependence of $r_m$,
\begin{equation}
    \frac{dr_m}{dt} = -\frac{f(r_m)}{2}\,.\label{drmdt}
\end{equation}
Since the past joint $r=r_m$ will be close to the horizon $r=r_+$ over time, taking the limit $r_m\to r_+$ is equivalent to the late time limit $t \to \infty$.

When another horizon (inner horizon) appears, the singularity becomes timelike instead. Therefore, the spacelike surface contributions from the intersection between the WDW patch and the singularity no longer exist. We have an additional future joint point named $r=r_{m}^1$ and the remaining past joint is now $r=r_{m}^2$ as in figure~\ref{double}. Similar to the one horizon case, by applying the Eddington-Finkelstein coordinates, we have the relation 
\begin{equation}
    \begin{aligned}
        &\frac{t}{2}+r^*(\infty)-r^*(r_m^1)=0\,,\\
        &\frac{t}{2}-r^*(\infty)+r^*(r_m^2)=0\,.
    \end{aligned}
\end{equation}
And the time dependence is
\begin{equation}\label{eq:rmt}
     \frac{dr_m^1}{dt} = \frac{f(r_m^1)}{2}\quad,\quad\frac{dr_m^2}{dt} = -\frac{f(r_m^2)}{2}\,.
\end{equation}
The late time limit, likewise, can be equivalently expressed as $r_m^2\to r_{+}$, $r_m^1\to r_{-}$ . In this case, we do not have a critical time as before, so the shape of the WDW patch is unchanged all the time. 

The full action to be evaluated in CA conjecture is \cite{Lehner:2016vdi}
\begin{align}
    S_{\text{WDW}} \quad =  \quad&\frac{1}{16\pi G_N}\int d^4x\sqrt{-g}\left(R-\frac12(\partial\phi)^2-V(\phi)-\frac14 e^{-\alpha\phi}F^2\right) \label{eq3.7}
   \\        &+\frac{1}{8\pi G_N}\int_\mathcal{B}d^3x \sqrt{-h}K-\frac{1}{8\pi G_N}\int_{\mathcal{B}'}d\lambda\,d^2\theta \sqrt{\gamma}\kappa\label{eq3.8}\\
   & +\frac{1}{8\pi G_N}\int_\Sigma d^2x \sqrt{\sigma}\eta + \frac{1}{8\pi G_N}\int_{\Sigma'} d^2x \sqrt{\sigma}a\label{eq3.9}\\
   &+\frac{1}{8\pi G_N}\int_{\mathcal{B}'}d\lambda\,d^2\theta \sqrt{\gamma}\Theta\log\left(l_{ct}\Theta\right)\label{eq3.10}\,.
\end{align}
The line~\eqref{eq3.7} is the bulk action terms. The line~(\ref{eq3.8}) is the codimension-one boundary contributions. The first term in (\ref{eq3.8}) is Gibbons-Hawking-York (GHY) surface term which comes from the spacelike surfaces at singularities as well as the timelike surfaces at the UV cutoff near the AdS boundary. The total timelike surface contribution will be time independent so we are free to ignore it in the calculation of time dependence of complexity in any case. The spacelike surface contributions at singularities are needed when there is only one horizon for our black holes. With two horizons, there is no such contribution from figure~\ref{double}. So the GHY boundary term is ignored for the black holes with two horizons. The second term in (\ref{eq3.8}) is the contribution from the null boundary of the WDW patch. From the discussion in \cite{Lehner:2016vdi}, we can affinely parametrize the generators and $\kappa$ can be set to zero. The line~(\ref{eq3.9}) is the contribution from the intersection of two boundaries. The first term is called the Hayward joint term, which represents the joint contribution that comes from the intersection of surfaces not null. This term plays no role in our discussion and we only include it here for the completion of the formula. The second term of (\ref{eq3.9}) is the joint term that comes from the intersection of WDW null boundaries with other surfaces. The contribution from joints at the singularities is zero, and the contribution from joints at the UV cutoff surface is time independent. The only useful term left is the joint contribution of two null WDW boundaries. The line~\eqref{eq3.10} is a counterterm introduced to protect the reparametrization invariance on the null boundaries, which has no effects on the variational principle.\footnote{In some particular cases like shock wave geometries, this kind of counterterm is needed to produce the expected results of complexity.} Based on these discussions, we can write the action that will be used in the following calculations:
\begin{equation}
    \begin{aligned}
             S_{\text{WDW}} \quad = \quad &\frac{1}{16\pi G_N}\int d^4x\sqrt{-g}\left(R-\frac12(\partial\phi)^2-V(\phi)-\frac14 e^{-\alpha\phi}F^2\right)\\
     &+\frac{1}{8\pi G_N}\int_\mathcal{B}d^3x \sqrt{-h}K+ \frac{1}{8\pi G_N}\int_{r_m} d^2x \sqrt{\sigma}a\\
     &+\frac{1}{8\pi G_N}\int_{\mathcal{B}'}d\lambda\,d^2\theta \sqrt{\gamma}\Theta\log\left(l_\text{ct}\Theta\right)\,.
    \end{aligned}
\end{equation}

A useful trick in the calculation is to divide the WDW patch symmetrically into left and right parts. These two parts can be further divided into three regions as figure~\ref{f1}. It is convenient to evaluate the $r$ integral in each part and finally multiply the result by a factor of 2 to get the full answer. 

\subsection{Complexity growth for neutral hyperbolic black hole}
In this section, we will calculate the time dependence of the holography complexity of the nontrivial neutral hyperbolic black hole. The neutral case is slightly simpler than the charged case so we choose to show the calculation in more detail first. From now on, we set $16\pi G_N = 1$ for simplicity.
\label{sec:3.1}

The bulk action is
\begin{equation}
    S_\text{bulk} = 2 V_2 \int\! drdt \, \sqrt{-g}\left(R-\frac12(\partial\phi)^2-V(\phi)\right),
\end{equation}
where $V_2$ is the area of the hyperbolic space $d\Sigma_2^2$, and we include a factor of 2 to account for the left part in figure~\ref{single}. We evaluate the $t$ integral in the Eddington-Finkelstein coordinates first and then we divide the $r$ integral into three parts as in figure~\ref{single},
\begin{equation}
    \begin{aligned}
        S_{\text{WDW}}^I &= 2V_2 \int^{r_+}_{r_m^1}dr \, \sqrt{-g}\left(R-\frac12(\partial\phi)^2-V(\phi)\right)\left(\frac{t}{2}+r^*_{\infty}-r^*(r)\right),\\
        S_{\text{WDW}}^{II} &= 2V_2 \int^{\infty}_{r_+}dr \, \sqrt{-g}\left(r^*_{\infty}-r^*(r)\right),\\
        S_{\text{WDW}}^{III} &= 2V_2 \int^{r_+}_{r_m^2}dr \, \sqrt{-g}\left(R-\frac12(\partial\phi)^2-V(\phi)\right)\left(-\frac{t}{2}+r^*_{\infty}-r^*(r)\right).
    \end{aligned}
\end{equation}
Then we obtain the time derivative on these parts and add them up,
\begin{equation}\label{eq:sbulk}
    \frac{dS_\text{bulk}}{dt} = -\frac{2 V_2}{(1+\alpha^2)}r^{\frac{3-\alpha^2}{1+\alpha^2}}\left(r-b\right)^{\frac{3\alpha^2-1}{1+\alpha^2}}\left(-b+r+r\alpha^2\right)\Bigg|_{r_m^1}^{r_m^2}\,.
\end{equation}
When there are two horizons, the points $r_m^1$ and $r_m^2$ are future and past intersections of the WDW null boundaries. When the geometry has only one horizon, the future meeting point $r_m^1$ will be replaced by the location of singularity $r=b$ for $b>0$ or $r=0$ for $b<0$. These two cases are related by a simple transformation $b\to-b$, $\alpha\to 1/\alpha$; therefore, we will only focus on the $b>0$ case in the following. From \eqref{eq:sbulk}, the contribution from the singularity $r=b$ has different values for $\alpha^2=1/3$ and $\alpha^2>1/3$,
\begin{equation}\label{3.15}
    \left\{
    \begin{aligned}
     -&\frac12 b^3\hspace{0.5cm}\alpha^2=\frac13\\
    &0\hspace{0.45cm}\hspace{0.45cm}\alpha^2>\frac13\,.
    \end{aligned}
\right.
\end{equation}
The next term is the GHY surface term. The action is
\begin{align}
    S_\text{surf} &= 4 V_2 \int\! dt \, \sqrt{-h} K\\
                &= \frac{2 V_2 }{ (1+\alpha^2)} \Bigl[b(3+\alpha^2)-4r(1+\alpha^2)+6(1-\frac{b}{r})^{\frac{3\alpha^2-1}{\alpha^2+1}}(r^3 (1+\alpha^2) -b)\Bigr]\notag\\
                &\hspace{0.5cm}\times\left(\frac{t}{2}+r_{\infty}^*-r^*(r)\right)\Bigg|_{r=b+\epsilon}.\notag
\end{align}
Here we introduce a regulator surface near the singularity $r=b$. Next we take the time derivative
\begin{equation}
    \frac{dS_\text{surf}}{dt}\Bigg|_{r=b+\epsilon} = \frac{V_2}{ (1+\alpha^2)}\Bigg(b(1+3\alpha^2)+4\epsilon(1+\alpha^2)
    -6\Bigl(\frac{\epsilon}{b+\epsilon}\Bigr)^{\frac{3\alpha^2-1}{\alpha^2+1}}(b+\epsilon)^2(b\alpha^2+\epsilon+\alpha^2\epsilon)\Bigg).
\end{equation}
Then we take the limit $\epsilon\to 0$,
\begin{equation}\label{eq:surfb}
    \frac{dS_\text{surf}}{dt}\Bigg|_{r=b} = \frac{V_2}{ (1+\alpha^2)}\Biggl(b (1+3\alpha^2)-6b^3\alpha^2\lim_{\epsilon\to0}\left(\frac{\epsilon}{b+\epsilon}\right)^{\frac{3\alpha^2-1}{\alpha^2+1}}\Biggr).
\end{equation}
We find that the limit in \eqref{eq:surfb} depends on the value of $\alpha$. For $0<\alpha^2<1/3$, a divergence appears; however, there is no such surface contribution because of the appearance of an additional horizon. For $\alpha^2\geq1/3$, the time derivative of the surface contribution at the singularity $r=b$ is
\begin{equation}
    \frac{dS_\text{surf}}{dt}\Bigg|_{r=b} = \left\{
    \begin{aligned}
    &\frac32V_2\left(b-b^3\right)\hspace{1.55cm}\alpha^2=\frac13\\
    &V_2\,b\left(\frac{1+3\alpha^2}{1+\alpha^2}\right) \hspace{1cm}\alpha^2>\frac13\,.
    \end{aligned}
    \right.
\end{equation}

The next contribution comes from the joint term. As we discussed before, the only meaningful contribution to the time dependence is the intersection between the null boundaries of the WDW patch. We have the formula \cite{Lehner:2016vdi}
\begin{equation}
    \begin{aligned}
    S_\text{joint} &= 2\int_{r_m} d^2x \sqrt{\sigma}a\\
              &= 2V_2 \sqrt{\sigma}\log\left(-\frac12 k\cdot \Bar{k}\right)\Bigg|_{r=r_m}\,,
    \end{aligned}
\end{equation}
where $a$, $k$, $\Bar{k}$ are defined according to \cite{Lehner:2016vdi}; $k$ and $\Bar{k}$ are the null normal vectors of the boundary given by
\begin{equation}
    \begin{aligned}
    k &= \beta\partial_{\mu}\left(t+r^*\right)\\
    \Bar{k} &= \Bar{\beta}\partial_{\mu}\left(t-r^*\right)\,.
\end{aligned}
\end{equation}
Then the joint contribution is 
\begin{equation}
S_\text{joint} = 2V_2\,r^2 \left(1-\frac{b}{r_m}\right)^{\frac{2\alpha^2}{1+\alpha^2}}\log\left(\frac{\beta\Bar{\beta}}{f(r_m)}\right).
\end{equation}
When $\alpha^2<1/3$, from figure~\ref{double}, there are two joint contributions at the future and past intersection of the null boundaries of the WDW patch. So the full contribution is the sum of these two contributions.  Recalling (\ref{eq:rmt}), we can obtain the time derivative of the total joint contribution
\begin{equation}
    \begin{aligned}
    \frac{dS_\text{joint}}{dt} &= \frac{V_2}{1+\alpha^2}\left[2r^2\left(1-\frac{b}{r}\right)^{\frac{3\alpha^2-1}{1+\alpha^2}}\left(-b+r+r\alpha^2\right)+b\left(\alpha^2-1\right)\right.\\
    &\left.\hspace{0.5cm}-2\left(1-\frac{b}{r}\right)^{\frac{\alpha^2-1}{\alpha^2+1}}\left(-b+r+r\alpha^2\right)f(r)\log\left(\frac{\beta^2}{f(r)}\right) \right]_{r_m^1}^{r_m^2}\,.
\end{aligned}
\end{equation}
Here we keep $f(r)$ implicitly in order to simplify our equation. When $\alpha^2\geq1/3$, the future joint is replaced by a spacelike surface. The joint contribution comes only from the past joint point $r=r_m^2$, and thus the time derivative is the expression above evaluated at $r=r_m^2$ only.

In the joint term, there are free coefficients $\beta$ and $\Bar{\beta}$ that reflect the ambiguity of complexity that comes from the null boundary of the WDW patch. To eliminate this ambiguity and, as \cite{Lehner:2016vdi} proposed, ensure the reparametrization invariance of the null boundary, we can introduce a counterterm that has no effect on the variational principle \cite{Goto:2018iay}
\begin{equation}
    S_\text{ct} = 2 \int\! d\lambda\,d^2\theta\sqrt{\gamma}\,\Theta\log\left(l_{ct}\Theta\right)\,.
\end{equation}
With an appropriate choice of affine parameter $\lambda$, we can evaluate the counterterms such that the combined contribution of counterterms and joint term can be independent of $\beta$ and $\Bar{\beta}$. Following \cite{Goto:2018iay}, a suitable choice is $\lambda=r/\beta$, where $\beta$ is exactly the constant that appeared before. The expansion $\Theta$ is
\begin{equation}
        \Theta= \partial_{\lambda}\log\sqrt{\gamma}
               =\frac{\beta\,\partial_r U(r)}{U(r)}
               = \frac{2\beta\left(-b+\alpha ^2 r+r\right)}{\left(\alpha ^2+1\right)(r-b)}\,.
\end{equation}
Then we can obtain the time derivative of the counterterm
\begin{equation}
    \begin{aligned}\label{eq:ct}
        \frac{dS_\text{ct}}{dt} &= \frac{4V_2}{1+\alpha^2}\left(1-\frac{b}{r}\right)^{\frac{\alpha ^2-1}{\alpha ^2+1}}\left(-b+r+r\alpha ^2\right)f(r)\log \left(\frac{2 \beta  l_{\text{ct}} \left(-b+\alpha ^2 r+r\right)}{r\left(\alpha ^2+1\right)\left(r-b\right)}\right)\Bigg|_{r=r_m}\,.
    \end{aligned}
\end{equation}
As we expected, the constant $\beta$ is eliminated by adding the counterterm into our full action. 

So far, we have prepared all the ingredients for the late time growth rate of complexity. First, we calculate the case of two horizons, 
\begin{align}\label{3.27}
    \frac{dS_\text{WDW}}{dt} &=\frac{dS_\text{bulk}}{dt}+\frac{dS_\text{joint}}{dt}+\frac{dS_\text{ct}}{dt}\\ 
                      &=V_2\left[b \left(\frac{\alpha^2-1}{\alpha^2+1}\right)-\frac{4\left(-b+r+r\alpha^2\right)}{1+\alpha^2}f(r)\left(1-\frac{b}{r}\right)^{\frac{-1+\alpha^2}{1+\alpha^2}}\log\left(\frac{U(r)}{l_{\text{ct}}\sqrt{f(r)}\partial_r U(r)}\right)\right]^{r_m^2}_{r_m^1}\notag\\
                      &\equiv\frac{dS_1}{dt}\Bigg|_{r_m^1}^{r_m^2}\,.\notag
\end{align}
Here we define a quantity $dS_1/dt$, which is simply a notation for future convenience. In the late time limit, the joint points $r_m^1$ and $r_m^2$ approach the inner horizon $r_-$ and outer horizon $r_+$, respectively. Additionally, $f(r)$ vanishes in the late time because $f(r_-)=f(r_+)=0$. So we obtain
\begin{equation}
    \begin{aligned}
    \lim_{t\to\infty}\frac{dS_\text{WDW}}{dt} &= V_2\, \left(\frac{\alpha^2-1}{\alpha^2+1}\right)b\,\Bigg|^{r_+}_{r_-}\\
                                         &= 0\,.
    \end{aligned}
\end{equation}
We find that the complexity growth of the neutral hyperbolic black holes vanishes in the late time period when there are two horizons. The same situation also occurs for the hyperbolic Schwarzchild-AdS black hole when its horizon radius is smaller than the AdS radius, which corresponds to a black hole with negative mass. We will discuss this further in section~\ref{sec:diss}.

As for the one horizon case, we have to add the GHY surface contribution and the bulk contribution at singularity based on the result of $0<\alpha^2<1/3$. Namely, when $\alpha^2\geq1/3$, 
\begin{equation}
    \begin{aligned}
        \frac{dS_\text{WDW}}{dt}\Bigg|_{\alpha^2\geq\frac13} &= \frac{dS_\text{bulk}}{dt}+\frac{dS_\text{surf}}{dt}+\frac{dS_\text{joint}}{dt}+\frac{dS_\text{ct}}{dt}\\
                            &=\left\{
                            \begin{aligned}
                                 &\frac{dS_1}{dt}\Bigg|_{r=r_m^2} + \frac12 V_2 b^3 + \frac32V_2\left(b-b^3\right)\hspace{1cm}\alpha^2=\frac13\\
                                &\frac{dS_1}{dt}\Bigg|_{r=r_m^2}+V_2 \left(\frac{1+3\alpha^2}{1+\alpha^2}\right)b\hspace{1.95cm}\alpha^2>\frac13\,.
                            \end{aligned}\right.\label{3.29}
    \end{aligned}
\end{equation}
Taking the late time limit gives
\begin{equation}
    \lim_{t\to\infty}\frac{dS_\text{WDW}}{dt}\Bigg|_{\alpha^2\geq\frac13} = 
    \left\{
     \begin{aligned}
     &V_2\left(b-b^3\right)\hspace{1.5cm}\alpha^2=\frac13\\
     &V_2\left(\frac{4b\alpha^2}{1+\alpha^2}\right)\hspace{1.1cm}\alpha^2>\frac13\,.
     \end{aligned}
    \right.
\end{equation}
Recall that the mass is given by 
\begin{equation}
    M = -2V_2\frac{1-\alpha^2}{1+\alpha^2}b\,.\notag
\end{equation}
Finally, the complexity growth rate in late time limit can be summarized as 
\begin{equation}\label{neutral}
    \lim_{t\to\infty}\frac{dS_\text{WDW}}{dt}\Bigg|_{b>0} = \left\{
    \begin{aligned}
    &\hspace{1.2cm}0\hspace{2.85cm}\alpha^2<\frac13\,,\\
    &2M+V_2\left(3b-b^3\right)\hspace{1.14cm}\alpha^2 = \frac13\,,\\
    &2M+V_2\left(\frac{4b}{1+\alpha^2}\right)\hspace{0.92cm}\alpha^2>\frac13\,.
    \end{aligned}
    \right.
\end{equation}

\subsection{Complexity for charged hyperbolic black hole}
\label{sec:3.2}
In this section, we compute the holographic complexity of charged hyperbolic black holes. Since many contents are similar to the neutral calculation in section \ref{sec:3.1}, we will omit the repeating detail. We begin with the case $b>0$. The bulk term is
\begin{equation}
    S_\text{bulk} = 2V_2\int \!drdt\sqrt{-g}\left(R-\frac12\left(\partial\phi\right)^2-V(\phi)-\frac14 e^{-\alpha\phi}F^2\right).
\end{equation}
As before, the factor of 2 means we have considered both sides of the WDW patch. Following the same procedure previously, we evaluate the $t$ integral and divide the $r$ integral into three parts. Then we take time derivatives of each term and add them up 
\begin{equation}
    \frac{dS_\text{bulk}}{dt} = 2V_2\left[\frac{q_e^2}{4r}+\left(r-b\right)^{\frac{3\alpha^2-1}{1+\alpha^2}}r^{\frac{3-\alpha^2}{1+\alpha^2}}\frac{\left(-b+r+r\alpha^2\right)}{1+\alpha^2}-\frac{c}{1+\alpha^2}\right]_{r_m^2}^{r_m^1}\,.
\end{equation}
Same as the discussion before, when there is only one horizon, the joint point $r_m^1$ is replaced by the location of singularity $r=b$ for $b>0$. The contribution at singularity $r=b$ is the same as \eqref{3.15}. 

The surface term is evaluated similarly to the neutral case. We take the time derivative and the limit $\epsilon\to0$,
\begin{equation}
    \frac{dS_\text{surf}}{dt}\Bigg|_{r=b} = \frac{V_2}{1+\alpha^2}\Biggl[(b+c)\left(1+3\alpha^2\right)-6b^3\alpha^2\lim_{\epsilon\to 0}\left(\frac{\epsilon}{b+\epsilon}\right)^{\frac{3\alpha^2-1}{1+\alpha^2}}\Biggr].
\end{equation}
Same as before, only when $\alpha^2\geq1/3$ will the surface term contribute,
\begin{equation}
    \frac{dS_\text{surf}}{dt} = \left\{
    \begin{aligned}
    &\frac32V_2\left(c+b-b^3\right), &\alpha^2=\frac13,\\
    &V_2\frac{1+3\alpha^2}{1+\alpha^2}\left(c+b\right),\quad &\alpha^2>\frac13.
    \end{aligned}
    \right.
\end{equation}
The joint term is
\begin{equation}
    \begin{aligned}
        \frac{dS_\text{joint}}{dt} &= \frac{V_2}{\left(\alpha ^2+1\right) (r-b)}\left[\left(\alpha ^2+1\right) r \left(c\left(1-\frac{b}{r}\right)+2r^3 \left(1-\frac{b}{r}\right)^{\frac{4\alpha ^2}{\alpha ^2+1}}\right)\right.\\
        &+b \left(-2c\left(1-\frac{b}{r}\right)-2r^3 \left(1-\frac{b}{r}\right)^{\frac{4\alpha ^2}{\alpha ^2+1}}+r\left(\alpha ^2-1\right)\left(1-\frac{b}{r}\right)\right)\\
        &\left.-2r\left(1-\frac{b}{r}\right)^{\frac{2\alpha ^2}{\alpha ^2+1}}f(r)\left(-b+r+r\alpha^2\right)\log{\left(\frac{\beta^2}{f(r)}\right)}\right]_{r_m^1}^{r_m^2}.
    \end{aligned}
\end{equation}
For the two horizons case, there are two joint contributions 
located at $r_m^1$ and $r_m^2$. For the one horizon case, the future joint $r_m^1$ will be replaced by the location of singularity $b$. 
The counterterm $S_{\text{ct}}$ is the same as the neutral case \eqref{eq:ct}. So far, we have all the ingredients to evaluate the full action of the WDW patch.

For the two horizons case, we obtain
\begin{align}\label{a<1/3full}
    \frac{dS_\text{WDW}}{dt}\Bigg|_{\alpha^2<\frac13} &= \frac{dS_\text{bulk}}{dt}+\frac{dS_\text{joint}}{dt}+\frac{dS_\text{ct}}{dt}\\
                        &= \Bigg[ V_2\left(-\frac{q_e^2}{r}-\frac{1-\alpha^2}{1+\alpha^2}\,b+\frac{3+\alpha^2}{1+\alpha^2}\,c\right)\notag\\
                        &-\frac{2V_2}{1+\alpha^2}\left(1-\frac{b}{r}\right)^{\frac{\alpha ^2-1}{\alpha ^2+1}}f(r)\left(-b+r+r\alpha^2\right)\log{\left(\frac{\beta^2}{f(r)}\right)}\notag\\
                        &\left.+\frac{4V_2}{1+\alpha^2}\left(1-\frac{b}{r}\right)^{\frac{\alpha ^2-1}{\alpha ^2+1}}f(r)\left(-b+r+r\alpha ^2\right)\log \left[\frac{2\beta l_{\text{ct}} \left(-b+\alpha ^2 r+r\right)}{r\left(\alpha ^2+1\right)\left(r-b\right)}\right]\notag\right]_{r_m^1}^{{r_m^2}}\notag\\
                        &=V_2 \left[\left(-\frac{q_e^2}{r}-\frac{1-\alpha^2}{1+\alpha^2}\,b+\frac{3+\alpha^2}{1+\alpha^2}\,c\right)\right.\notag\\
                        &\left.-\frac{4}{1+\alpha^2}\left(1-\frac{b}{r}\right)^{\frac{\alpha ^2-1}{\alpha ^2+1}}f(r)\left(-b+r+r\alpha ^2\right)\log\left(\frac{U(r)}{l_{\text{ct}}\sqrt{f(r)}\partial_r U(r)}\right)\right]_{r_m^1}^{{r_m^2}}\notag\\ 
                        &\equiv\frac{dS_2}{dt}\Bigg|_{r_m^1}^{r_m^2}\notag\,.
\end{align}
Likewise, the term $dS_2/dt$ is a notation for future convenience. In the late time limit, 
\begin{equation}
    \begin{aligned}
        \lim_{t\to\infty}\frac{dS_\text{WDW}}{dt}&= V_2\left[-\frac{q_e^2}{r}-\frac{1-\alpha^2}{1+\alpha^2}\,b+\frac{3+\alpha^2}{1+\alpha^2}\,c\right]\Bigg|_{r_{-}}^{r_{+}}\\
                                            &=q_e^2V_2\left(\frac{1}{r_{-}}-\frac{1}{r_{+}}\right)\,.
    \end{aligned}
\end{equation}
For the one horizon case, we obtain
\begin{equation}
    \frac{dS_\text{WDW}}{dt}\Bigg|_{\alpha^2\geq\frac13} = \left\{
    \begin{aligned}
        &\frac{dS_2}{dt}\Bigg|_{r=r_m^2}  +\frac12 V_2\,b^3+ \frac32V_2\left(c+b-b^3\right) \hspace{1cm} \alpha^2=\frac13\\
        &\frac{dS_2}{dt}\Bigg|_{r=r_m^2}  +V_2\frac{1+3\alpha^2}{1+\alpha^2}\left(c+b\right) \hspace{2.3cm} \alpha^2>\frac13\,.
    \end{aligned}
    \right.\label{3.39}
\end{equation}
In the late time limit, 
\begin{equation}\label{3.40}
    \begin{aligned}
        \lim_{t\to\infty}\frac{dS_\text{WDW}}{dt}\Bigg|_{b>0} &= \lim_{t\to\infty}\left(\frac{dS_\text{bulk}}{dt}+\frac{dS_\text{surf}}{dt}+\frac{dS_\text{joint}}{dt}+\frac{dS_\text{ct}}{dt}\right)\\
                                             &= \left\{
                                             \begin{aligned}
                                                 &V_2\left(-\frac{q_e^2}{r_+}+4c+b-b^3\right)\hspace{1.3cm}\alpha^2=\frac13\\
                                                 &V_2\left(-\frac{q_e^2}{r_+}+\frac{4\alpha^2}{1+\alpha^2}\,b+4c\right)\hspace{0.88cm}\alpha^2>\frac13
                                             \end{aligned}
                                             \right.\\
                                             &= \left\{
                                             \begin{aligned}
                                                 &2M-V_2\left(\frac{q_e^2}{r_+}-3b+b^3\right)\hspace{1.15cm}\alpha^2=\frac13\\
                                                 &2M-V_2\left(\frac{q_e^2}{r_+}-\frac{4}{1+\alpha^2}\,b\right)\hspace{0.9cm}\alpha^2>\frac13\,.
                                             \end{aligned}
                                             \right.
    \end{aligned}
\end{equation}
In the last equality, we introduce the mass of charged hyperbolic black hole by setting $k=-1$ in \eqref{massgen},
\begin{equation}\label{mass}
    M = 2V_2\left(c-\frac{1-\alpha^2}{1+\alpha^2}\,b\right).
\end{equation}
Finally, we collect the results,
\begin{equation}\label{4dcharged}
    \lim_{t\to\infty}\frac{dS_\text{WDW}}{dt}\Bigg|_{b>0} = \left\{
    \begin{aligned}
        &V_2\left[q_e^2\left(\frac{1}{r_-}-\frac{1}{r_+}\right)\right], &\alpha^2<\frac13\,,\\
        &2M-V_2\left(\frac{q_e^2}{r_+}-3b+b^3\right), &\alpha^2=\frac13\,,\\
        &2M-V_2\left(\frac{q_e^2}{r_+}-\frac{4}{1+\alpha^2}\,b\right),\quad &\alpha^2>\frac13\,.
    \end{aligned}
    \right.
\end{equation}
Remember that we only discussed the case with positive $b$ above. When $b<0$, as appendix~\ref{sec:noh} shows, the black hole will always have two horizons for any $\alpha$. So the late time growth of holographic complexity with negative $b$ for all $\alpha$ is
\begin{equation}\label{b<0}
    \lim_{t\to\infty}\frac{dS_\text{WDW}}{dt}\Bigg|_{b<0} = V_2\left[q_e^2\left(\frac{1}{r_-}-\frac{1}{r_+}\right)\right]\,.
\end{equation}

We expect the late time growth rate of the holographic complexity to be non-negative. When $\alpha^2<1/3$ and $\alpha^2>1/3$, it is obvious to show that the results are positive. When $\alpha^2=1/3$, we have
\begin{equation}
    \begin{aligned}
        -\frac{q_e^2}{r_+}+4c+b-b^3 &=4c\left(1-\frac{3b}{4r_+}\right)+b-b^3\\
        & >c+b-b^3\\
        & =-r_+ + r_+^3+b-b^3\\
        & =\left(r_+-b\right)\left(r_+^2+br_++b^2-1\right)\\
        & >\left(r_+-b\right)\left(b+b^2\right)>0\,.
    \end{aligned}
\end{equation}
In the second line, we use the condition that the horizon radius must be outside the singularity, namely, $r_+>b$. By applying \eqref{c_of_rh} with $\alpha^2=1/3$ and $k=-1$, we obtain the equation $c=-r_++r_+^3$. We use this relation to obtain the third equality. When $c>0$, the equation $c=-r_++r_+^3$ only has one positive solution with $r_+>1$. We apply this constraint to obtain the last equality. These calculations ensure that our result is positive as we expect. 
\section{Conclusion and discussion}
\label{sec:diss}
\subsection{The late time dependence of holographic complexity for general $k$}
It is intriguing to compare our result with its spherical counterpart from \cite{Goto:2018iay}, 
\begin{equation}
    \lim_{t\to\infty}\frac{dS_\text{WDW}}{dt} = \left\{
    \begin{aligned}
    &4\pi\left[q_e^2\left(\frac{1}{r_-}-\frac{1}{r_+}\right)\right], &\alpha^2<\frac13\,,\\
    &2M-4\pi\left(\frac{q_e^2}{r_+}+3b+b^3\right), &\alpha^2=\frac13\,,\\
    &2M-4\pi\left(\frac{q_e^2}{r_+}+\frac{4}{1+\alpha^2}\,b\right),\quad &\alpha^2>\frac13\,.
    \end{aligned}
    \right.
\end{equation}
Obviously, there are similarities between these two solutions and the planar solution listed in appendix~\ref{sec:ap}. Note that $b>0$ is required for the spherical and planar black hole solutions. With positive $b$, we collect all these solutions as
\begin{equation}
\begin{boxed}
    {\lim_{t\to\infty}\frac{dS_\text{WDW}}{dt} = \left\{
    \begin{aligned}
        &V_{2,k}\left[q_e^2\left(\frac{1}{r_-}-\frac{1}{r_+}\right)\right], &\alpha^2<\frac13\,,\\
        &2M-V_{2,k}\left(\frac{q_e^2}{r_+}+3kb+b^3\right), &\alpha^2=\frac13\,,\\
        &2M-V_{2,k}\left(\frac{q_e^2}{r_+}+\frac{4k}{1+\alpha^2}\,b\right),\quad &\alpha^2>\frac13\,.
    \end{aligned}
    \right.}
\end{boxed}
\label{eq:master}
\end{equation}
With $k=-1$, $0$, $1$, this unified result reduces to hyperbolic, planar, and spherical cases correspondingly. It is interesting to discuss the neutral limit of these cases. We begin with the limit $b\to0^+$, which is applicable to all different $k$. In this limit, we obtain AdS-Schwarzchild black holes with mass $2V_2\,c$. When $\alpha^2\geq 1/3$, the causal structure is kept under the limit $b\to 0^+$ so we can directly set $b=0$ in the late time result \eqref{eq:master} and we can obtain the expected result $2M$. When $\alpha^2<1/3$, the causal structure will change under the limit $b\to 0^+$ (from two horizons to one horizon) so we cannot directly apply this limit in the late time result \eqref{eq:master}. We have to properly include the contribution at singularity and then we can obtain the expected result $2M$ as in \cite{Brown:2015lvg, Carmi:2017jqz}. 

Recall that the hyperbolic black hole $(k=-1)$ is special because $b$ and $c$ can be negative in this case. When $b<0$, $c<0$, there are always two horizons for any $\alpha$ and the late time growth rate of holographic complexity is \eqref{b<0}. Now if we take the limit $b\to 0^-$, we obtain a hyperbolic AdS-Schwarzchild black hole with negative mass (because $c<0$). Therefore, the resulting black hole will also have two horizons \cite{Carmi:2017jqz}. This causal structure similarity allows us to directly set $b=0$ in \eqref{b<0} and we obtain the expected vanishing result.

Another neutral limit, $c=0$, is only applicable to the hyperbolic case $k=-1$, which is also a distinctiveness of the hyperbolic black holes. When $b>0$, since the causal structure is the same between the charged and the neutral case $(c=0)$, we can set $c=0$ in \eqref{4dcharged} and the result agrees with what we obtain in \ref{sec:3.1}. When $b<0$, the causal structure of the neutral black hole $(c=0)$ can be obtained from the $b>0$ case by applying the transformation $b\to-b$, $\alpha\to1/\alpha$, namely, two horizons when $\alpha^2>3$ and one horizon when $\alpha^2\leq3$. Hence we have to properly include the contribution at singularity instead of directly set $c=0$ in \eqref{4dcharged}. The above discussion about neutral limit emphasizes the essential position of causal structure in the study of holographic complexity. 

An important feature of this neutral hyperbolic black hole is the existence of phase transition \cite{Ren:2019lgw}. As a consequence, the R\'{e}nyi entropy also experiences a phase transition \cite{Bai:2022obp}. Hence, it is natural to expect that the holographic complexity also has a phase transition when the temperature changes. From~\eqref{tands}, the temperature depends on the horizon radius $r_+$. When $\alpha^2\neq1/3$, we have 
\begin{equation}\label{tinrh}
    T=\frac{f'(r_+)}{4\pi}=\frac{1}{4\pi(1+\alpha^2)}\left[(3-\alpha^2)r_+^{\frac{1+\alpha^2}{1-3\alpha^2}}-(1-3\alpha^2)r_+^{-\frac{1+\alpha^2}{1-3\alpha^2}}\right].
\end{equation}
When $\alpha^2=1/3$, we have $T=\sqrt{1-b}/2\pi$. When $\alpha^2=3$, we have $T=\sqrt{1+b}/2\pi$. When $1/3<\alpha^2<3$, there is a minimal temperature above zero, 
\begin{equation}
    T_{\text{m}}=\frac{\sqrt{(3-\alpha^2)(3\alpha^2-1)}}{2\pi(1+\alpha^2)}\,, 
\end{equation}
above which there are two different black holes at a given temperature. The parameter $b$ can be expressed in terms of $r_+$ by applying the equation $f(r_+)=0$, which may have one or two real solutions $r_{+,\pm}$. We have
\begin{align}\label{binrh}
    b&=r_{+,\pm} - r_{+,\pm}^{\frac{3-\alpha^2}{1-3\alpha^2}}\nonumber\\
    &=\Biggl(\frac{\pm\sqrt{4 \pi ^2 \left(\alpha^2+1\right)^2 T^2+\left(3-\alpha^2\right) \left(1-3 \alpha^2\right)}+2 \pi  \left(\alpha^2+1\right) T}{3-\alpha^2}\Biggr)^{\frac{1-3 \alpha^2}{\alpha^2+1}}\nonumber\\
    &\hspace{0.5cm}-\Biggl(\frac{\pm\sqrt{4 \pi ^2 \left(\alpha^2+1\right)^2 T^2+\left(3-\alpha^2\right) \left(1-3 \alpha^2\right)}+2 \pi  \left(\alpha^2+1\right) T}{3-\alpha^2}\Biggr)^{\frac{3- \alpha^2}{\alpha^2+1}}.
\end{align}
To study the phase transition between the scalar hairy black hole and the SAdS black hole, we also need to write the mass of the SAdS black hole in terms of temperature, which is
\begin{equation}\label{MinT}
        M = 2V_2 c=\frac{4V_2}{27} \left(\sqrt{4 \pi ^2 T^2+3}+2 \pi  T\right) \left(2 \pi T\sqrt{4 \pi ^2 T^2+3}-3+4\pi^2 T^2\right).
\end{equation}
When $\alpha^2\leq1/3$ and $\alpha^2\geq3$, the temperature reaches zero at
\begin{equation}
    r_+=\left(\frac{1-3\alpha^2}{3-\alpha^2}\right)^{\frac{1-3\alpha^2}{2(1+\alpha^2)}}, \qquad b=\frac{2(1+\alpha^2)}{3-\alpha^2}\left(\frac{1-3\alpha^2}{3-\alpha^2}\right)^{\frac{1-3\alpha^2}{2(1+\alpha^2)}}.
\end{equation}

When $\alpha^2\leq1/3$, the scalar hairy black hole has a horizon at $r_{+,+}$. As for $b=0$, the corresponding temperature is $T_0=1/2\pi$ which is the phase transition point between the SAdS black hole and the hairy hyperbolic black hole. On the left side of this transition point, the parameter $b$ is positive. When $\alpha^2<1/3, b>0$, the late time complexity growth rate of the scalar hairy black hole is zero according to \eqref{neutral}. The late time complexity growth rate of the SAdS black hole is $2M$. We notice that the growth rate is continuous at $T_0$ but its derivative is not. So there is a first-order phase transition for the late time growth rate. When $\alpha^2=1/3, b>0$, the late time complexity growth rate of the scalar hairy black hole is $V_2(b-b^3)$ according to \eqref{neutral}. By expressing $b$ in terms of $T$, we can also find that there is also a first-order phase transition. When $\alpha^2\geq3$, the lower temperature side will correspond to negative $b$. This case, as we said before, is similar to the $\alpha^2<1/3$ case under a transformation $\alpha\to1/\alpha, b\to-b$.

\begin{figure}[]
\centering
\subfigure[]{\includegraphics[width=0.4\textwidth]{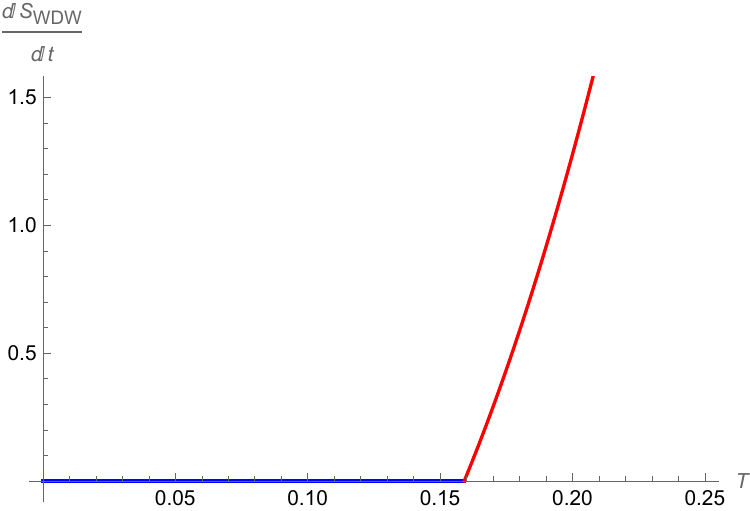}}\label{phase1}
\subfigure[] {\includegraphics[width=0.4\textwidth]{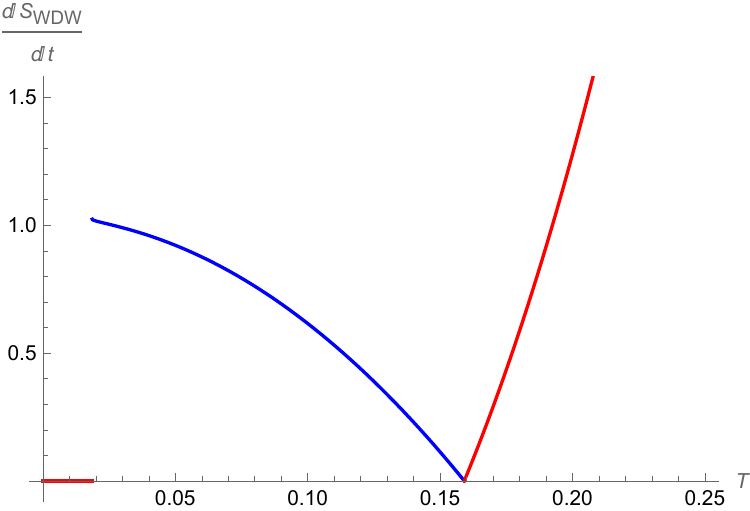}}\label{phase2}
\subfigure[]{\includegraphics[width=0.4\textwidth]{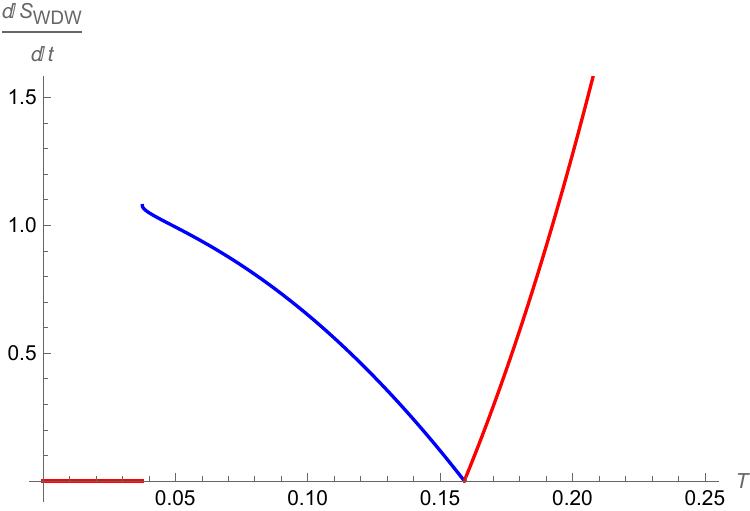}}\label{phase3}
\subfigure[]{\includegraphics[width=0.4\textwidth]{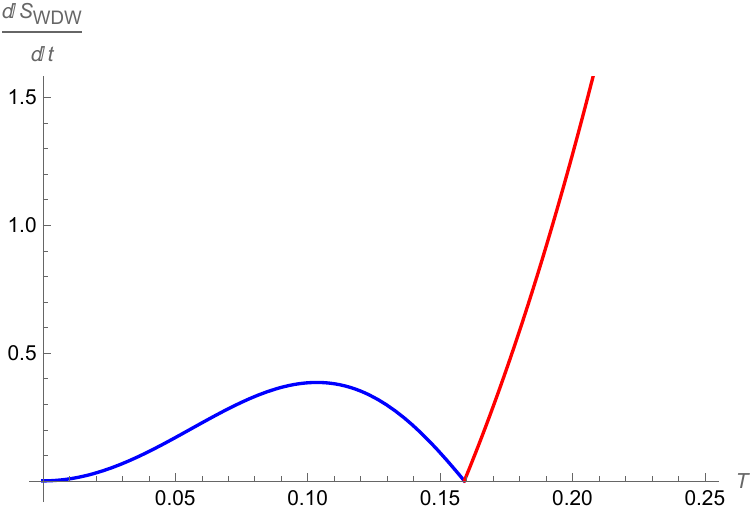}}\label{phase4}
\caption{(a) $\alpha^2<1/3$ and $\alpha^2>3$, (b) $1/3<\alpha^2<1$, (c) $1<\alpha^2<3$, (d) $\alpha^2=1/3$ and $\alpha^2=3$. The blue and red curves correspond to the holographic complexity of scalar hairy black hole and SAdS black hole, respectively.}
\label{figphase}
\end{figure}

When $1/3<\alpha^2<3$, there is a nonzero minimal temperature $T_\text{m}$ at 
\begin{equation}
    r_+=\left(\frac{3\alpha^2-1}{3-\alpha^2}\right)^{\frac{1-3\alpha^2}{2(1+\alpha^2)}},\qquad b=\frac{4(1-\alpha^2)}{3-\alpha^2}\left(\frac{3\alpha^2-1}{3-\alpha^2}\right)^{\frac{1-3\alpha^2}{2(1+\alpha^2)}}.
\end{equation}
When $1/3<\alpha^2<1$, the stable solution corresponds to the horizon $r_{+,+}$ and $b$ is positive on the lower temperature side. When $b>0$, the late time growth rate of the hairy black hole is $ 4V_2b\alpha^2/(1+\alpha^2)$ and the late time growth rate of SAdS is $2M$. We then have
\begin{equation}
   8\pi V_2= \frac{d(2M)}{dT}\quad\neq\quad\frac{d}{dT}\left(\frac{dS_{\text{WDW}}}{dt}\right)=8\pi V_2\frac{\left(3-\alpha^2\right)^{\frac{3 \alpha^2-1}{\alpha^2+1}} \left(3 \alpha^2-1\right)^{\frac{2-2 \alpha^2}{\alpha^2+1}}}{\alpha^2-1} \,.
\end{equation}
So there is a first-order phase transition at $T_0$. When $1<\alpha^2<3$, the stable solution corresponds to the horizon $r_{+,-}$ and $b$ is negative on the lower temperature side. 
When $b<0$, the late time growth rate of the hairy black hole is $ -4V_2b/(1+\alpha^2)$. Still, by making a transformation $\alpha\to1/\alpha, b\to-b$ we will recover the case $1/3<\alpha^2<1$.
So there is also a first-order phase transition at $T_0$. In conclusion, we find that the holographic complexity does experience a first-order phase transition as in figure~\ref{figphase}.  

\subsection{Full time dependence of holographic complexity}
\subsubsection{Charged case ($c\neq0$)}
\begin{figure}[H]
\centering
\begin{minipage}[t]{0.4\textwidth}
\centering
\includegraphics[scale=0.45]{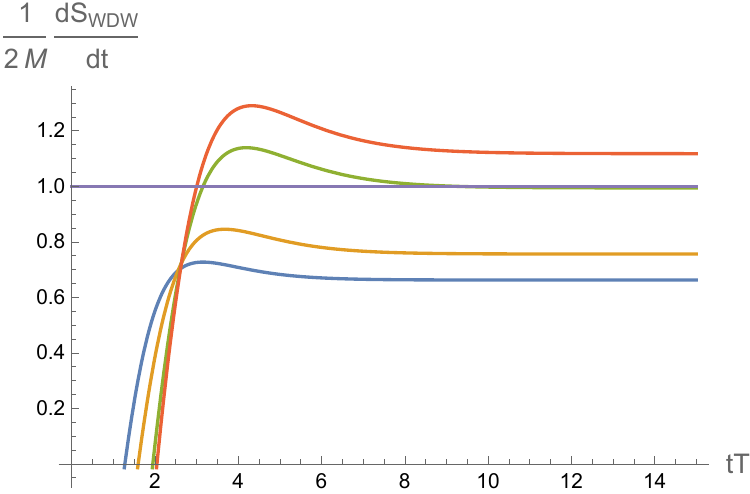}
\caption{Full time behavior for different parameters when $\alpha^2=1/4$, $b=0.5$. The blue, yellow, green, and red curves correspond to $c=1.2$, $c=0.8$, $c=0.55$, $c=0.5$, respectively. The purple line represents the ratio $\frac{1}{2M}\frac{dS_{\text{WDW}}}{dt}=1$.}
\label{bound_pic1}
\end{minipage}
\hspace{10mm}
\begin{minipage}[t]{0.4\textwidth}
\centering
\includegraphics[scale=0.45]{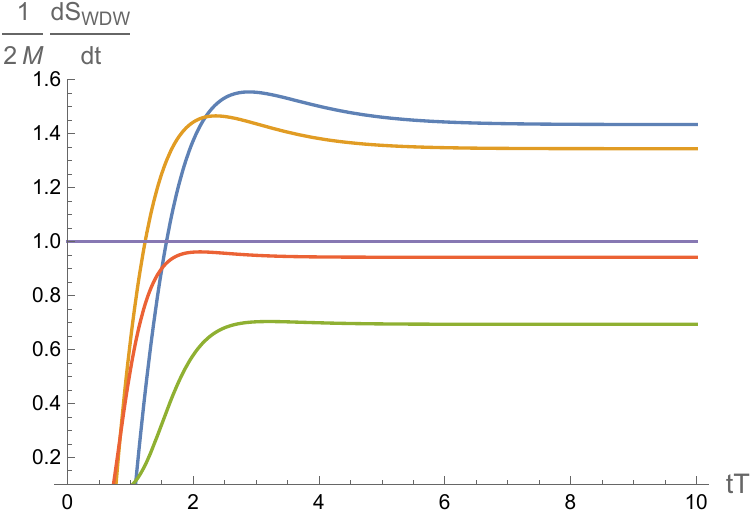}
\caption{Full time behavior for different parameters when $\alpha^2=1/4, b<0$. The blue curve corresponds to $b=-1, c=-0.1$. The yellow curve corresponds to $b=-1, c=-0.01$. The green curve corresponds to $b=-10, c=-0.3$. The red curve corresponds to $b=-10, c=-0.4$. The purple line represents the ratio $\frac{1}{2M}\frac{dS_{\text{WDW}}}{dt}=1$.}
\label{bound_pic2}
\end{minipage}
\end{figure}

In this section, we discuss the full time dependence of the complexity and compare our result with the Lloyd bound. The full time dependence when $c\neq 0$ has been shown in \eqref{a<1/3full} and \eqref{3.39}. Notice that the mass of the hyperbolic black hole can be negative. Obviously, the Lloyd bound is always violated when mass is negative. When the mass is positive, we numerically show the full time dependence of the holographic complexity growth rate for various parameters. From figure~\ref{bound_pic1}, when $\alpha^2<1/3$, $b>0$, the Lloyd bound is violated not only in early time but also in late time for some parameters. For $b<0$, as in the discussion above, there are also two horizons and the result of the growth rate of complexity is similar to the case $\alpha^2<1/3$, $b>0$. From figure~\ref{bound_pic2}, we also notice that the bound violation is also possible in the whole time period.

\begin{figure}[b]
\centering
\subfigure[]{\includegraphics[width=0.4\textwidth]{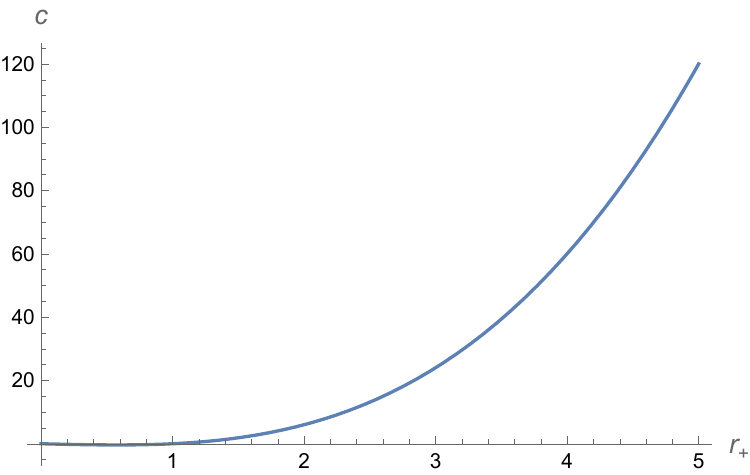}\label{figcr1}}
\subfigure[] {\includegraphics[width=0.4\textwidth]{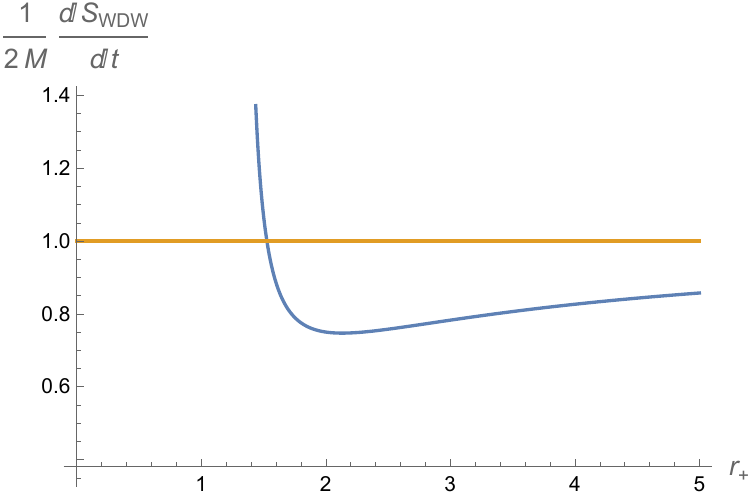}\label{figcr2}}
\caption{(a) Shape of \eqref{4.3} when $\alpha^2=1/3$, $b=1$. (b) Shape of \eqref{4.4} when $\alpha^2=1/3$, $b=1$.}
\label{fig5}
\end{figure}

\begin{figure}[H]
    \centering
    \includegraphics[width=0.6\textwidth]{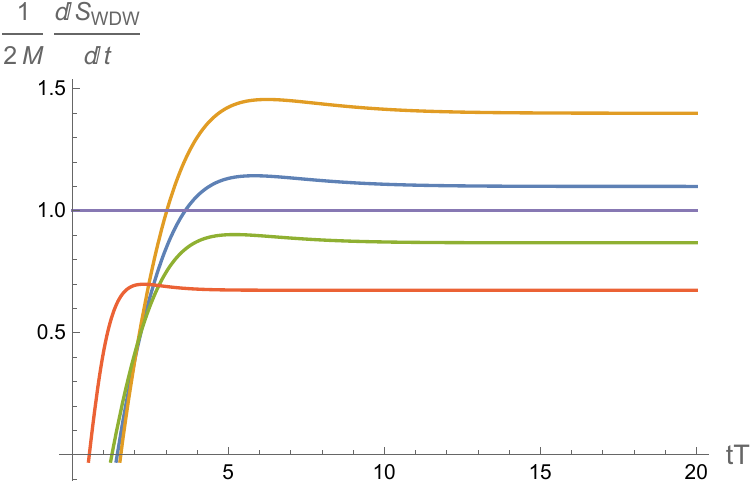}
    \caption{Full time behavior for different parameters when $\alpha^2=1/3$, $b=1$. The curves from top to bottom correspond to $c=0.7$, $c=0.8$, $c=1$, and $c=5$, respectively. The purple line represents the ratio $\frac{1}{2M}\frac{dS_{\text{WDW}}}{dt}=1$.}
    \label{bound_pic3}
\end{figure}

When $\alpha\geq1/3$, there is only one horizon for positive $b$, $c$. This simpler structure benefits our further analysis of the parameter space. First, we consider the case $\alpha^2=1/3$. Recall \eqref{3.39}, 
  \begin{equation}
      \lim_{t\to\infty}\frac{dS_{\text{WDW}}}{dt}=V_2\left(-\frac{q_e^2}{r_+}+4c+b-b^3\right)\notag\,.
  \end{equation}
From equation $f(r_+)=0$, we can express $c$ as
\begin{equation}\label{4.3}
    c=-r_++r_+^3\left(1-\frac{b}{r_+}\right)^{\frac{3\alpha^2-1}{1+\alpha^2}}\,.
\end{equation}
Together with the mass formula \eqref{mass}, we have 
\begin{equation}\label{4.4}
    \lim_{t\to\infty}\frac{1}{2M}\frac{dS_{\text{WDW}}}{dt}= \frac{\left(\frac{-3b}{r_+}+4\right)\left(-r_++r_+^3\right)+b-b^3}{4\left(-r_++r_+^3-\frac12 b\right)}\,.
\end{equation}
Let this equal $1$ and solve $r_+$,
\begin{equation}
    r_+ = \sqrt{\frac{6-b^2}{3}}\,.
\end{equation}
Remember that the horizon must locate outside the singularity. So when $b>0$,
\begin{equation}\label{4.6}
    r_+ = \sqrt{\frac{6-b^2}{3}}>b \quad\Longrightarrow\quad 0<b<\frac{\sqrt{6}}{2}\,.
\end{equation}
From figure \ref{fig5}, we can conclude that, for some fixed $b$ in $0<b<\sqrt{6}/2$, the Lloyd bound is violated in late time when $b<r_+<\sqrt{6-b^2}/\sqrt{3}$. We can express the range of $c$ as 
\begin{equation}
  \max\left(b^3-b,0\right)< c <\sqrt{\frac{6-b^2}{3}}\left(\frac{3-b^2}{3}\right)\,.
\end{equation}
We express the lower bound of $c$ as $\max\left(b^3-b,0\right)$ because we only consider positive $b$ and $c$. And the term $b^3-b$ is negative when $0<b<1$.  From figure \ref{bound_pic3}, we find exactly this behavior in the full time dependence. 
When $\alpha^2>1/3$, although we cannot obtain an explicit condition for the bound violation, we can still make some progress based on the explicit result of \eqref{eq:master}. From \eqref{3.40}, the late time dependence is given by
\begin{equation}
    \lim_{t\to\infty}\frac{dS_{\text{WDW}}}{dt}=V_2\left(-\frac{q_e^2}{r_+}+4c+\frac{4\alpha^2}{1+\alpha^2}b\right)\,.
\end{equation}
Similarly, we consider the quantity
\begin{equation}\label{4.8}
     \lim_{t\to\infty}\frac{1}{2M}\frac{dS_{\text{WDW}}}{dt}= \frac{\left(\frac{-4b}{\left(1+\alpha^2\right)r_+}+4\right) c+\frac{4\alpha^2}{1+\alpha^2}b}{4\left(c-\frac{1-\alpha^2}{1+\alpha^2} b\right)}=\frac{c-\frac{\frac{c}{r_+}-\alpha^2}{1+\alpha^2}b}{c-\frac{1-\alpha^2}{1+\alpha^2} b}\,.
\end{equation}
Let this equal $1$, we obtain $c=r_+$. From \eqref{4.8}, we conclude the following:
\begin{itemize}
    \item When $0<b<r_+<c$, \eqref{4.8} will be greater than $1$, which means that the Lloyd bound is violated with the corresponding $b$ and $c$ in the late time limit.  
    \item When $0<b<c<r_+$, \eqref{4.8} will be smaller than $1$, which means that the Lloyd bound is satisfied with the corresponding $b$ and $c$ in the late time limit.
    \item When $0<c<b<r_+$, \eqref{4.8} will be greater than $1$, which means that the Lloyd bound is violated with the corresponding $b$ and $c$ in the late time limit. 
\end{itemize}

\begin{figure}[H]
\centering
\begin{minipage}[t]{0.4\textwidth}
\centering
\includegraphics[scale=0.45]{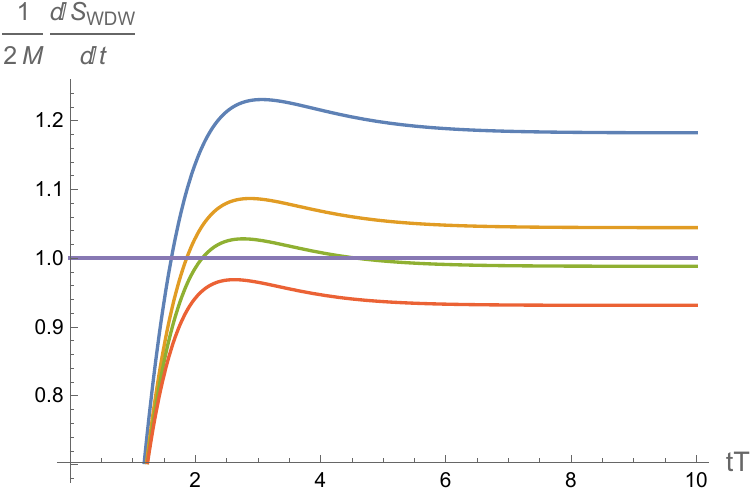}
\caption{Full time behavior for different parameters when $\alpha^2=1/2$, $b=1$. The blue, yellow, green, and red curves correspond to $c=1.2$, $c=1.5$, $c=1.7$, and $c=1.8$, respectively. The purple line represents the ratio $\frac{1}{2M}\frac{dS_{\text{WDW}}}{dt}=1$.}
\label{bound_pic4}
\end{minipage}
\hspace{10mm}
\begin{minipage}[t]{0.4\textwidth}
\centering
\includegraphics[scale=0.45]{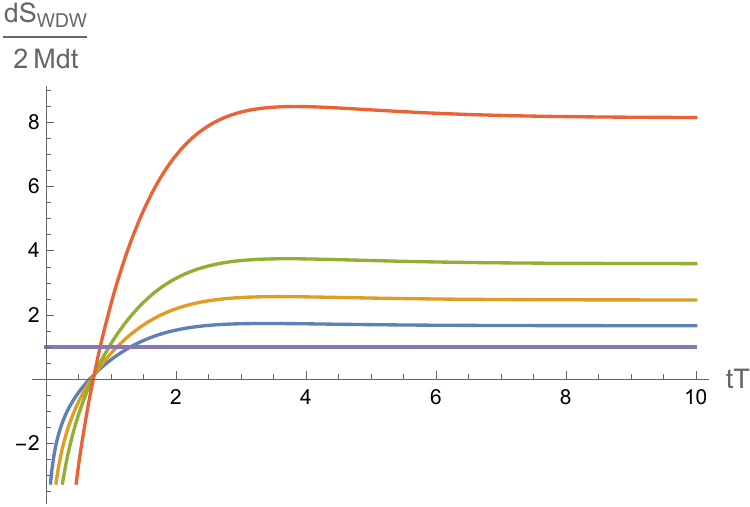}
\caption{Full time behavior for different parameters when $\alpha^2=1/2, b=1$.  The blue, yellow, green, and red curves correspond to $c=0.8$, $c=0.6$, $c=0.5$, and $c=0.4$, respectively. The purple line represents the ratio $\frac{1}{2M}\frac{dS_{\text{WDW}}}{dt}=1$.}
\label{bound_pic6}
\end{minipage}
\end{figure}

For example, we set $b=1$, $\alpha^2=1/2$. From \eqref{4.3}, we can get $c=r_+\approx 1.65$. In figure \ref{bound_pic4}, we find the full time dependence supports our analysis before. We find that the late time value is approached from above so the Lloyd bound may be violated in early time even if it is obeyed in the late time limit. In figure \ref{bound_pic6}, we find that when $0<c<b$, the Lloyd bound is always violated.

\subsubsection{Neutral case ($c=0$)}
\begin{figure}[H]
    \centering
    \includegraphics[scale=0.7]{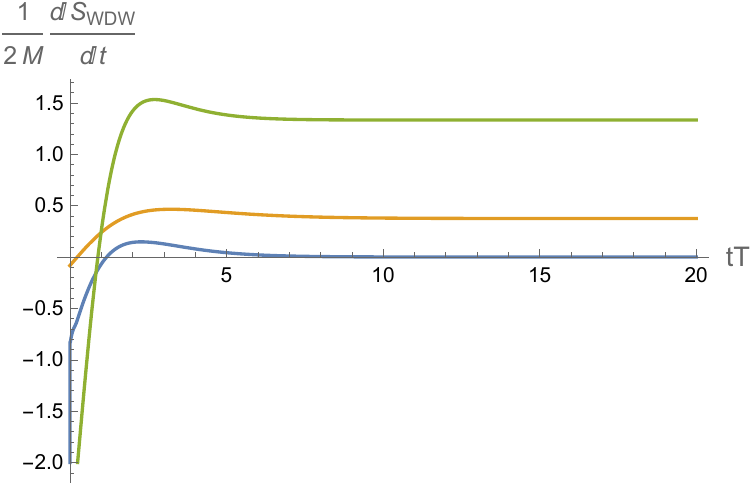}
    \caption{Full time dependence when $c=0$, $b=0.5$. From top to bottom, the curves correspond to $\alpha^2=2$, $\alpha^2=1/3$, $\alpha^2=1/4$.}
    \label{bound_pic5}
\end{figure}

In the neutral case, the full time dependence has been shown in \eqref{3.27} and \eqref{3.29}. The late time growth rate is 
\begin{equation}\label{neutral4}
    \lim_{t\to\infty}\frac{dS_\text{WDW}}{dt}\Bigg|_{b>0} = \left\{
    \begin{aligned}
    &\hspace{1.2cm}0\hspace{2.85cm}\alpha^2<\frac13\,,\\
    &2M+V_2\left(3b-b^3\right)\hspace{1.14cm}\alpha^2 = \frac13\,,\\
    &2M+V_2\left(\frac{4b}{1+\alpha^2}\right)\hspace{0.92cm}\alpha^2>\frac13\,.
    \end{aligned}
    \right.
\end{equation}
When $\alpha^2<1$, $b>0$, the mass of the black hole is always negative. So the Lloyd bound is always violated by the positive growth rate. When $\alpha^2<1/3$, we find that the growth rate increases in the early time and finally approaches zero from above, which matches the late time result. When $\alpha^2\geq1/3$, the full time dependence is similar. When $\alpha^2>1$, the mass is positive but the Lloyd bound is still violated from \eqref{neutral4}. See figure~\ref{bound_pic5}.

The violation of the Lloyd bound has been studied in various models, including hyperscaling violating geometries \cite{Swingle:2017zcd,Alishahiha:2018tep} and Einstein-scalar theories \cite{Mahapatra:2018gig}. This neutral hyperbolic black hole with $\alpha^2\geq1/3$ can be regarded as another example that supports \cite{Mahapatra:2018gig} even though these two theories have different scalar potentials. Additionally, the $\alpha^2<1/3$ case is interesting because of the vanishing late time growth rate. The vanishing growth rate also appears in different contexts such as RN-AdS black hole \cite{Goto:2018iay}, small hyperbolic SAdS black hole \cite{Carmi:2017jqz}, and Jackiw-Teitelboim (JT) gravity \cite{Goto:2018iay, Alishahiha:2018swh,Brown:2018bms}. In the first two cases, there is no good reason to not accept this vanishing. However, in JT gravity, this vanishing growth rate is physically unacceptable because of the conjectured duality between JT gravity and the SYK model. In the latter, the system is supposed to be highly chaotic, so the quantum complexity is expected to grow linearly even after an extremely long time. Back to our case, an interesting observation is that the boundary of the hyperbolic spacetime is conformal to a de Sitter space \cite{Marolf:2010tg}, which implies that the holographic dual theory is a quantum field theory in de Sitter space. Recently, some people suggest that the holographic dual of de Sitter space may be a double-scale limit Sachdev-Ye-Kitaev (SYK) model \cite{Susskind:2022dfz,Lin:2022nss,Susskind:2022bia,Rahman:2022jsf,Susskind:2023hnj}, which experiences hyperfast scrambling. The vanishing complexity growth rate seems to be inconsistent with this conjecture.

\vspace{6pt}
In our result, we find that the dilaton field can accelerate the growth rate of holographic complexity for the hyperbolic black hole. This accelerated effect leads to the violation of the Lloyd bound in the late time limit. In previous work
\cite{An:2018xhv,Cai:2017sjv}, authors studied the holographic complexity of a spherical EMD system with $k=1$, $\alpha^2=1$. The late time growth rate is $2M-\mu Q-D$ and the dilaton field contribution $D$ appears to slow down the growth of complexity. Despite lacking rigorous proof, many arguments have been suggested to explain this effect. In \cite{Zhang:2017nth}, authors related this slowing effect to the breaking of conformal symmetry. In \cite{Ghodrati:2018hss}, it is explained that the dilaton field actually describes how strongly the open strings couple to one another. In other words, the coupling constant is decided by the expectation value of the dilaton field. The state with many interacting strings will be hard to move to another state so the complexity rate will be lower. The accelerated effect found in this article may alert us to reconsider the effect of dilaton on holographic complexity. 



There are still many other topics that deserve further discussion. An interesting feature of the hyperbolic model is that the boundary is conformal to a de Sitter spacetime. An interesting direction is to discuss the complexity of quantum field theory in de Sitter spacetime and compare it with the bulk computation in this article. The boundary side may provide more evidence for a possible explanation for the abnormal behavior mentioned above. Additionally, we only apply CA conjecture in this article. It can be interesting to apply other conjectures such as CV conjecture and CV 2.0 and compare their result to our CA calculation.


\acknowledgments
J.R. thanks Shubho R. Roy for helpful discussions. This work is supported in part by the National Natural Science Foundation of China (NSFC) under Grant No.~11905298, and Fundamental Research Funds for the Central Universities, Sun Yat-sen University under Grant No.~23qnpy61.

\appendix
\section{Number of horizons}\label{sec:noh}
For the hyperbolic black holes, the number of horizons is either one or two.

(A) $b>0$: the singularity is at $r=b$. Consider the function
\begin{equation}
f_1(r)=\biggl(-1-\frac{c}{r}\biggr)\biggl(1-\frac{b}{r}\biggr)^\frac{1-3\alpha^2}{1+\alpha^2}+\frac{r^2}{L^2},
\end{equation}
It is obvious that the $\alpha^2=1/3$ is special because the exponent will be zero. We begin our analysis with this situation. The function $f_1(r)$ will be
\begin{equation}
    f_1(r) = \frac{r^2}{L^2}-1-\frac{c}{r}\,.
\end{equation}
With $b>0$, $c>0$, this function always has one zero. That is, we will always have one horizon when $\alpha^2=1/3$, $b>0$. For the cases $\alpha^2\neq 1/3$, we analyze the behavior of the function near $r=b$ and $r=0$, and then we can have an approximate shape of the function that gives us information about the number of zeros of $f_1(r)$, namely, the number of horizons. 

\begin{figure}[H]
    \centering
    \subfigure[]{\includegraphics[width=0.49\textwidth]{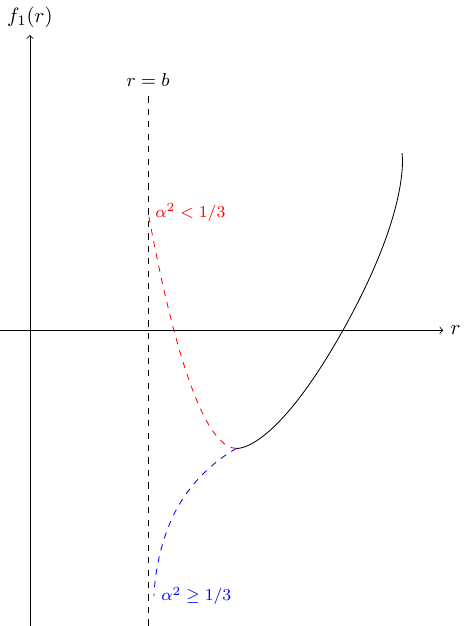}}
    \subfigure[]{\includegraphics[width=0.49\textwidth]{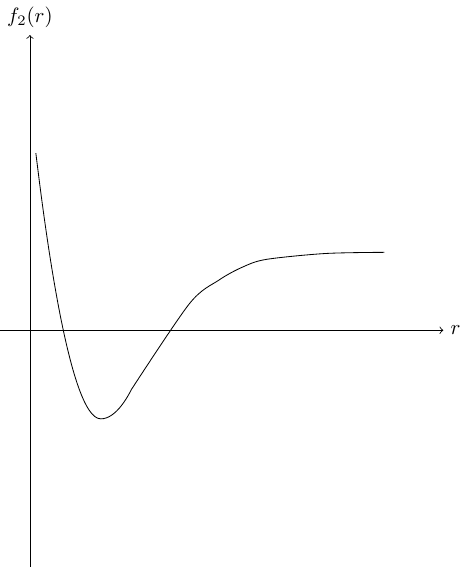}}
    \caption{(a) Schematic plot of $f_1(r)$ for different $\alpha$ when $b>0$. (b) Schematic plot of $f_2(r)$ for any $\alpha$ when $b<0$.}
    \label{fig:no-hor}
\end{figure}

The behavior near $r=b$ is
\begin{equation}
f_1(r)\to\begin{cases}
b^2/L^2>0,\quad &\alpha^2<1/3\\
-\infty, &\alpha^2>1/3\,.
\end{cases}
\end{equation}
The behavior near infinity is
\begin{equation}
    f_1(r)\to\infty\,.
\end{equation}
The derivative of $f_1(r)$ is
\begin{equation}
    f'_1(r) = 2r+\frac{-b(2c+r)+b(2c+3r)\alpha^2+cr(1+\alpha^2)}{r^3(1+\alpha^2)}\biggl(1-\frac{b}{r}\biggr)^\frac{-4\alpha^2}{1+\alpha^2}.
\end{equation}
Its behavior near $r=b$ is
\begin{equation}
f'_1(r)\to\begin{cases}
-\infty,\quad &\alpha^2<1/3\\
\infty, &\alpha^2>1/3\,.
\end{cases}
\end{equation}
The behavior near infinity is 
\begin{equation}
    f'_1(r)\to\infty\,.
\end{equation}
Based on these analyses, we can draw the function $f_1(r)$ approximately as figure \ref{fig:no-hor}(a).

(B) $b<0$: the singularity is at $r=0$. Consider the function
\begin{equation}
f_2(r)=\frac{1}{r^2}\biggl(-1-\frac{c}{r}\biggr)\biggl(1-\frac{b}{r}\biggr)^\frac{1-3\alpha^2}{1+\alpha^2}+\frac{1}{L^2}.
\end{equation}
The behavior near $r=0$ is
\begin{equation}
f_2(r)\to \frac{|c|\cdot|b|^{\frac{1-3\alpha^2}{1+\alpha^2}}}{r^{\frac{4}{1+\alpha^2}}}\to\infty,
\end{equation}
provided $c\neq 0$. The behavior near infinity is
\begin{equation}
    f_2(r)\to \frac{1}{L^2}>0\,.
\end{equation}
The derivative of $f_2(r)$ is
\begin{equation}
f'_2(r)=\frac{-4bc+br(-3+\alpha^2)+r(3c+2r)(1+\alpha^2)}{r^5(1+\alpha^2)}\biggl(1-\frac{b}{r}\biggr)^\frac{-4\alpha^2}{1+\alpha^2}.
\end{equation}
Its behavior near $r=0$ is
\begin{equation}
    f'_2(r)\to-\infty\,.
\end{equation}
The behavior near infinity is
\begin{equation}
    f'_2(r)\to0
\end{equation}
With this information, we can draw the function $f_2(r)$ approximately as figure \ref{fig:no-hor}.

In conclusion, for $b>0$ and $c\geq 0$, the hyperbolic black holes have two horizons when $\alpha^2<1/3$, one horizon when $\alpha^2\geq 1/3$; for $b<0$ and $c<0$, the hyperbolic black holes always have two horizons. For $b<0$ and $c=0$, there is one horizon when $\alpha^2\leq 3$ and two horizons when $\alpha^2>3$.

\section{Complexity growth of planar EMD black holes}
\label{sec:ap}
This appendix will briefly discuss the holographic complexity of planar EMD black holes. The charged version of this black hole is similar to the other two sisters. But the neutral case $(c=0)$ is a little more special. Recall the solution of the metric for the EMD system is
\begin{equation}
    ds^2 =-f(r)\tsp dt^2+\frac{1}{f(r)}\tsp dr^2+U(r)\tsp d\tsp\Sigma_{2,k}^2\,.
\end{equation}
In the planar case, namely, $k=0$, 
\begin{equation}
    \begin{aligned}
        f&=-\frac{c}{r}\left(1-\frac{b}{r}\right)^{\frac{1-\alpha^2}{1+\alpha^2}}+r^2\left(1-\frac{b}{r}\right)^{\frac{2\alpha^2}{1+\alpha^2}}\,,\\
        U&=r^2\left(1-\frac{b}{r}\right)^{\frac{2\alpha^2}{1+\alpha^2}}\,.
    \end{aligned}
\end{equation}
In the charged case, the causal structure for these solutions is the same as the discussion before. There will be two horizons when $0<\alpha^2<1/3$ and one horizon when $\alpha^2>1/3$. Then we can similarly evaluate the WDW action as before. Start with the bulk term, 
\begin{equation}
        S_\text{bulk}=2V_{\,2}^{k=0}\int\!dr\,\left(R-2\left(\partial\phi\right)^2-V(\phi)-\frac14e^{-\alpha\phi}F^2\right)\left(\frac{t}{2}+r^*_{\infty}-r^*(r)\right)
\end{equation}
The time derivative, 
\begin{equation}
    \frac{dS_\text{bulk}}{dt}=\frac{2\,V_{\,2}^{k=0}}{1+\alpha^2}\left[c\left(1-\frac{b}{r}\right)-\left(1-\frac{b}{r}\right)^{\frac{3\alpha^2-1}{1+\alpha^2}}r^2\left(-b+r+r\alpha^2\right)\right]^{r_m^2}_{r_m^1}\,.
\end{equation}
The surface term at the regular surface near the singularity $r=b$, 
\begin{equation}
    \begin{aligned}
        S_\text{surf}\Bigg|_{r=b+\epsilon} = &\frac{2V_{\,2}^{k=0}}{\left(1+\alpha^2\right)\left(b+\epsilon\right)}\Bigg[c\left(3\epsilon\left(\alpha^2+1\right) +b\left(3\alpha^2+1\right)\right)\\
        &-6\epsilon^3\left(\frac{\epsilon }{b+\epsilon }\right)^{-\frac{4}{\alpha ^2+1}}\left(\alpha ^2(b+\epsilon )+\epsilon \right)\Bigg]\left(\frac{t}{2}+r^*_{\infty}-r^*(r)\right)\,.
    \end{aligned}
\end{equation} 
The time derivative and limit $\epsilon\to\infty$, 
\begin{equation}
    \frac{dS_\text{surf}}{dt} = \left\{
    \begin{aligned}
        &\frac32\left(c-b^3\right)\hspace{1.2cm}\alpha^2=\frac13\\
        &\frac{1+3\alpha^2}{1+\alpha^2}\,c\hspace{1.35cm}\alpha^2>\frac13\,.
    \end{aligned}
    \right.
\end{equation} 
Then comes the joint term,
\begin{equation}
    S_\text{joint} = r^2\left(1-\frac{b}{r}\right)^{\frac{2\alpha^2}{1+\alpha^2}}\log\left[\frac{1}{f(r)}\right]\,.
\end{equation}
Still, we make $f(r)$ implicitly in order to waive the complex formula. As usual, the time derivative is
\begin{equation}
    \begin{aligned}
        \frac{dS_\text{joint}}{dt} &=V_{\,2}^{k=0}\Bigg[-\frac{q_e^2}{2r}+\frac{c}{1+\alpha^2}+2r^2\left(r-\frac{b}{1+\alpha^2}\right)\left(1-\frac{b}{r}\right)^{\frac{3 \alpha ^2-1}{\alpha ^2+1}}\\
        &\hspace{0.5 cm}-2\left(\frac{-b+\alpha ^2 r+r}{1+\alpha^2}\right)f(r)\log\left[\frac{1}{f(r)}\right]\Bigg]\,.
    \end{aligned}
\end{equation}
As it always is, the counterterm $S_{ct}$ is still invariant. At last, we can get the full WDW action growth in late time, 
\begin{equation}
    \lim_{t\to\infty}\frac{dS_\text{WDW}}{dt} = \left\{
    \begin{aligned}
         &V_{\,2}^{k=0}\,q_e^2\left(\frac{1}{r_-}-\frac{1}{r_+}\right)\hspace{1cm}\alpha^2<\frac13\\
         &2M-V_{\,2}^{k=0}\left(\frac{q_e^2}{r_+}+b^3\right)\hspace{0.6cm}\alpha^2=\frac13\\
         &2M-V_{\,2}^{k=0}\,\frac{q_e^2}{r_+}\hspace{2.05cm}\alpha^2>\frac13\,.
    \end{aligned}
    \right.
\end{equation}
In the neutral case, if we take $b=0$, then the solutions will reduce to the planar Schwarzchild black holes and its holographic complexity will simply be $2M$. Another neutral limit $c=0$ is irregular, i.e., the horizon approaches to the singularity located at $r=b$. In other words, we are left with a naked singularity instead of a black hole. The complexity of this naked singularity has recently been calculated in \cite{Katoch:2023dfh}.

In the neutral limit $c=0$, the growth rate of the complexity is zero except for the $\alpha^2=1/3$ case. The complexity of the neutral limit $c=0$ was obtained to be constant in \cite{Katoch:2023dfh}. The small inconsistency at $\alpha^2=1/3$ needs further investigation. 

\section{The holographic complexity in five dimensions}\label{sec:hc5d}
In this appendix, we briefly show the holographic complexity of hyperbolic EMD black holes in AdS$_5$. The action is
\begin{equation}
    S=\int d^5x\sqrt{-g}\left(R-\frac14 e^{-\alpha\phi}F^2-\frac12(\partial\phi)^2-V(\phi)\right),
\end{equation}
where $F=dA$ as (\ref{eq:action4}). The potential of the dilaton field of this AdS$_5$ bulk spacetime is
\begin{equation}
   V(\phi)=-\frac{12}{(4+3\alpha^2)^2L^2}\Bigl[3\alpha^2(3\alpha^2-2)e^{-4\phi/3\alpha}
+36\alpha^2e^{\frac{3\alpha^2-4}{6\alpha}\phi}+2(8-3\alpha^2)e^{\alpha\phi}\Bigr].\label{eq:potential5}
\end{equation}
As $\phi\to0$, $V(\phi)\to-12/L^2-(2/L^2)\phi^2+\cdots$, where the first term is the cosmological constant as in the four-dimension case. We set $L=1$ as before. The solution of the fields are 
\begin{align}
    ds^2&=-f(r)dt^2+\frac{1}{g(r)}dr^2+U(r)d\Sigma^2_3\,,\\
    A&=2\sqrt{\frac{3b^2c^2}{4+3\alpha^2}}\left(\frac{1}{r_+^2}-\frac{1}{r^2}\right)dt\quad,\quad e^{\alpha\phi}=\left(1-\frac{b^2}{r^2}\right)^{\frac{6\alpha^2}{4+3\alpha^2}}\,,
\end{align}
where
\begin{align}
    f(r)&=\left(-1-\frac{c^2}{r^2}\right)\left(1-\frac{b^2}{r^2}\right)^{\frac{4-3\alpha^2}{4+3\alpha^2}}+r^2\left(1-\frac{b^2}{r^2}\right)^{\frac{3\alpha^2}{4+3\alpha^2}}\,,\\
    g(r)&=f(r)\left(1-\frac{b^2}{r^2}\right)^{\frac{3\alpha^2}{4+3\alpha^2}}\quad,\quad U(r)=r^2\left(1-\frac{b^2}{r^2}\right)^{\frac{3\alpha^2}{4+3\alpha^2}}\,,
\end{align}
where $b$ and $c$ are parameters as before. The mass is 
\begin{equation}
    M=3V_3\left(c^2-\frac{4-3\alpha^2}{4+3\alpha^2}b^2\right)\,.
\end{equation}
The chemical potential $\mu$ and charge density $q_e$ are
\begin{equation}
\mu=2\sqrt{\frac{3b^2c^2}{4+3\alpha^2}}\,\frac{1}{r_+^2},\qquad q_e=4\sqrt{\frac{3b^2c^2}{4+3\alpha^2}}\,.
\end{equation}
The first law of thermodynamics is satisfied: $d\varepsilon=Tds+\mu dq_e$.

To calculate the holographic complexity, we have to analyze the causal structure. When $0<\alpha<2/\sqrt{6}$, there will be two horizons in our spacetime. Same as before, we only consider the nonextremal solutions with $c^2>b^2>0$. When $\alpha\geq2/\sqrt{6}$, we have only one horizon and a spacelike singularity at $r=b$. Since the calculation is similar to the four-dimension case, we simply list the results as follows:
\begin{equation}
    \lim_{t\to\infty}\frac{dS_\text{WDW}}{dt} = \left\{
    \begin{aligned}
         &V_{\,3}\,\mu q_e\left(\frac{r_+^2}{r_-^2}-1\right), &\alpha^2<\frac23\\
         &2M-V_{\,3}\left(\mu q_e-4b^2+2b^4\right), &\alpha^2=\frac23\\
         &2M-V_{\,3}\,\left(\mu q_e-\frac{24}{4+3\alpha^2}b^2\right),\quad &\alpha^2>\frac23
    \end{aligned}
    \right.
\end{equation}
Comparing with the four-dimension case \eqref{4dcharged}, the structure of the results is similar. This result may enable us to obtain the expression in any dimension.

\end{document}